\documentclass{article}

\usepackage{myspconf}
\usepackage{amsmath,amssymb,amsthm}
\usepackage{graphicx}
\usepackage{overpic}
\usepackage{caption}
\usepackage{subcaption}
\usepackage{float}
\usepackage{bm}
\usepackage{algorithm}
\usepackage{algorithmicx}
\usepackage{algpseudocode}

\usepackage{indentfirst}
\usepackage{color}

\usepackage{epstopdf}
\usepackage{grffile}
\usepackage{booktabs}
\usepackage{url}

\usepackage{tikz}
\usetikzlibrary{spy,decorations.fractals}
\usetikzlibrary{lindenmayersystems}
\usetikzlibrary{shadings}
\usepackage{pgf}
\usetikzlibrary{patterns,backgrounds,calc,arrows.meta}
\pgfdeclarelayer{foreground}
\pgfsetlayers{background,main,foreground}

\makeatletter
\renewcommand*\section{\@startsection {section}{1}{\z@}%
	{-2.4ex \@plus -1ex \@minus -.2ex}%
	{1.3ex \@plus.2ex}%
	{}}
\renewcommand*\subsection{\@startsection {subsection}{2}{\z@}%
	{-1.8ex \@plus -.5ex \@minus -.2ex}% before
	{1ex \@plus.2ex}% after
	{}}
\makeatother
\DeclareMathOperator*{\argmin}{argmin} % no space, limits underneath in displays

\newcommand{\x}		{\mathbf{x}}

\newcommand{\y}		{\mathbf{y}}
\newcommand{\A}		{\mathbf{A}}
\newcommand{\z}	 {\mathbf{z}}
\newcommand{\W}	 {\mathbf{W}}

\newcommand{\R}	     {\mathsf{R}}

\newcommand*{\cR}{\mathcal{R}}
\newcommand*{\cD}{\mathcal{D}}

\newcommand*{\mb}[1]{{\mathbf{#1}}}
\newcommand*{\bbE}{\mathbb{E}}

\newcommand{\BLUE}{\textcolor{black}}

\newcommand*{\rth}{\mathrm{th}}

\newcommand{\nin}{\noindent}
\def\R#1{(\ref{#1})}

\newcommand{\bes}[1]{\begin{equation*} #1 \end{equation*}}

\newcommand{\eas}[1]{\begin{align*} #1 \end{align*}}

\title{Momentum-Net for Low-Dose CT Image Reconstruction}
\name{Siqi Ye$^1$, Yong Long*$^1$, Il Yong Chun$^2$ \thanks{This work is supported by NSFC (61501292). *Yong Long (email: yong.long@sjtu.edu.cn). \textregistered 20XX IEEE. Personal use of this material is permitted. Permission from IEEE must be obtained for all other uses, in any current or future media, including reprinting/republishing this material for advertising or promotional purposes, creating new collective works, for resale or redistribution to servers or lists, or reuse of any copyrighted component of this work in other works. }}
\address{$^1$University of Michigan - Shanghai Jiao Tong University Joint Institute,\\
	Shanghai Jiao Tong University, Shanghai, China\\
	$^2$Department of Electrical Engineering, University of Hawai'i at Manoa, HI, USA}

\begin{document}
	\maketitle
\begin{abstract}
This paper applies the recent fast iterative neural network framework, Momentum-Net, using appropriate models to low-dose X-ray computed tomography (LDCT) image reconstruction. At each layer of the proposed Momentum-Net, the model-based image reconstruction module solves the majorized penalized weighted least-square problem, and the image refining module uses a four-layer convolutional neural network (CNN). Experimental results with the NIH AAPM-Mayo Clinic Low Dose CT Grand Challenge dataset show that the proposed Momentum-Net architecture significantly improves image reconstruction accuracy, compared to a state-of-the-art \emph{noniterative} image denoising deep neural network (NN), WavResNet (in LDCT). We also investigated the spectral normalization technique that applies to image refining NN learning to satisfy the nonexpansive NN property; however, experimental results show that this does not improve the image reconstruction performance of Momentum-Net.
\end{abstract}
\section{Introduction}
\label{sec:intro} 
X-ray computed tomography (CT) plays an important role in diverse clinical applications; however, radiation exposure to patients is of great concern. Low-dose CT (LDCT) is a major approach to resolve the radiation issue.

Conventional analytical CT image reconstruction methods such as filtered back-projection (FBP) \cite{feldkamp1984practical} often provide unsatisfactory results in LDCT. The model-based image reconstruction (MBIR) methods enable better reconstruction qualities in low-dose scans, by incorporating the CT imaging physics, the measurement noise statistics, and appropriate prior models of the unknown object \cite{fessler2000statistical,Polyenergetic02,survey2013-ct}. Applying the learned regularizers to MBIR has taken researchers one step closer to enabling LDCT in clinics \cite{xu:12:ldx,pwls-ultra2018,Chun&Fessler:18TIP,Chun&Fessler:20TIP,Chun&David:19TSPL}.
 Although MBIR methods using learned regularizers gave promising results in LDCT, their algorithmic convergence speed is not yet fully optimized \cite{Chun&Fessler:20TIP}.
% the computational cost is expansive even with some fast iterative optimization algorithms, e.g., the relaxed LALM agorithm\cite{nien:16:rla} and FISTA \cite{Beck&Teboulle:09SIAM}.

%and characterized by fewer parameters, thereby avoiding over-fitting and generalizing well to unseen data especially when training samples are limited.
Researchers also have explored image denoising deep neural network (NN) methods that 
remove noise in coarsely reconstructed LDCT image via good mapping capabilities of deep NNs.
Examples include the residual encoder-decoder convolutional neural network (RED-CNN) \cite{red-cnn17}, and the wavelet residual network (WavResNet) that learns a mapping of contourlet coefficients between low- and standard-dose image pairs \cite{WavResNet18}. The image denoising performance of such deep NNs largely depends on the quality of reference training images \cite{Zheng&etal:19TCI} and the number of training pairs \cite{Lim&etal:19TMI}. In CT imaging, however, due to risks of high radiation exposure to patients, collecting many high-quality reference images is challenging in practice.
%It is even impossible to collect exactly paired low- and high-dose images. 
When training samples are limited, NNs with high model complexity, e.g., deep NNs, can suffer from high overfitting risks \cite{Zheng&etal:19TCI, Chun&etal:19MICCAI, Lim&etal:19TMI}. 
The iterative neural network (INN) approach can moderate the overfitting issue by using both image refining (or denoising) NNs with low to moderate complexity, and MBIR optimization in an iterative fashion.
%One way to \BLUE{moderate} these problems is the \BLUE{P}lug-and-\BLUE{P}lay (PnP) strategy \BLUE{that} integrates some \BLUE{image refining (or denoising) NNs that 
%appropriately approximate the true unknown image} into an iterative MBIR algorithm,. 
Recently, \cite{Chun&Fessler:18IVMSP} constructed the \emph{BCD-Net} architecture by generalizing a block coordinate descent (BCD) algorithm that solves the MBIR problem using convolutional regularizers. Incorporating refined images via layer-wise refining NNs to MBIR modules, BCD-Net showed better generalization capability and reconstruction quality than a recent (\emph{noniterative}) denoising deep NN, FBPConvNet \cite{fbpconvnet17}, and a state-of-the-art INN method, ADMM-Net \cite{Yang&etal:16NIPS, Chan&Wang&Elgendy:17TCI}, respectively \cite{Chun&etal:19MICCAI}.
A practical limitation of applying BCD-Net (and ADMM-Net) to LDCT is that the MBIR module at each layer needs multiple inner iterations, leading to substantial increase of the total image reconstruction time. To resolve this issue, \cite{chun2019momentumnet} proposed a fast INN architecture, \emph{Momentum-Net}, that is constructed by generalizing a block proximal extrapolated gradient MBIR algorithm using convolutional regularizers. Different from BCD-Net, Momentum-Net introduces additional extrapolation modules and uses practical closed-form MBIR solutions, which can accelerate the LDCT image reconstruction process.

This paper applies Momentum-Net to LDCT image reconstruction with appropriate model selections. At each Momentum-Net layer, we use the majorized penalized weighted least-square (PWLS) MBIR cost and a four-layer residual convolutional neural network (CNN), so-called SimpleCNN \cite{ryu2019plug}. Our experiments show that the proposed Momentum-Net architecture significantly improves LDCT image reconstruction accuracy over a state-of-the-art \emph{noniterative} image denoising deep NN, WavResNet \cite{WavResNet18}. In addition, we investigate the performance of applying Real Spectral Normalization (RSN) \cite{ryu2019plug} to image refining NN learning; however, this normalized refining NN learning does not improve the reconstruction accuracy of Momentum-Net.

\section{Methods}\label{sec:formulation}
\begin{figure}[!t]
%\begin{figure}[!htp]
%	\vspace{-0.1in}
	\centering
	\includegraphics[width=1\linewidth]{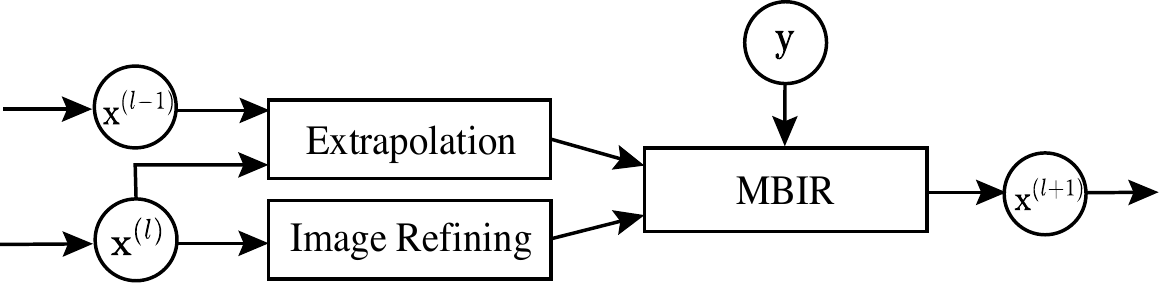}
   \vspace{-1pc}
	\caption{Architecture of Momentum-Net at the $l$th layer \cite{chun2019momentumnet}, for $l  \geq 0$.  $\x^{(l+1)}$ represents the reconstructed image at the $l$th Momentum-Net layer. The initial inputs $\x^{(0)} = \x^{(-1)}$.}
	\label{fig:arch}
	\vspace{-0.75pc}
\end{figure}

The objective of CT image reconstruction is to reconstruct the unknown linear attenuation map $\x \in \mathbb{R}^{N_p}$ of an object from the post-log sinogram $\y \in \mathbb{R}^{N_d}$.   
In the Momentum-Net architecture,each layer consists of three core modules: image refining, extrapolation, and MBIR. The abstract Momentum-Net architecture is shown in Fig.~\ref{fig:arch}. 
We then describe each module in detail.

\subsection{Image refining module}\label{sec:refining}
We design the image refining module to provide a high-quality prior estimate for the subsequent MBIR module. In different Momentum-Net layers, we learn a different four-layer residual CNN, SimpleCNN \cite{ryu2019plug}.
The denoised image at the $l\rth$ layer is obtained by the $l\rth$ layer SimpleCNN denoiser $\cD_{\theta^{(l+1)}}$ defined as follows:
\begin{equation}
\label{eq:denoiser}
\cD_{\theta^{(l+1)}}(\x^{(l)}) = \x^{(l)} - \cR_{\theta^{(l+1)}}(\x^{(l)}).
%\vspace{-0.05in}
\end{equation}
where $\cR_{\theta^{(l+1)}}$ denotes learned residual mapping at the $l\rth$ layer and $\theta^{(l+1)}$ represents its parameter set.
%The learned residual mapping at the $l\rth$ layer is denoted as $\cR_{\theta^{(l+1)}}$, where $\theta^{(l+1)}$ represents the network parameters; 
%\RED{the denoised image at the }
%the denoised image at the $l\rth$ layer $\cD_{\theta^{(l+1)}}(\cdot)$ is obtained by

Next, we investigate the effect of learning nonexpansive $\{ \cR_{\theta^{(l+1)}}(\cdot) : l \geq 0 \}$ via the real spectral normalization (RSN) technique \cite{ryu2019plug}, on INN performance. We refer to this learned residual denoiser with RSN as \emph{SimpleCNN-RSN}. As nonexpansive residual mapping $\cR_{\theta^{(l+1)}}(\cdot)$ does not guarantee nonexpansive denoiser $\cD_{\theta^{(l+1)}}(\cdot)$ in \R{eq:denoiser}, 
we remove the skip connection from denoiser \R{eq:denoiser}, and investigate the effect of RSN-based learning on INN performance.
We refer to this learned denoiser with RSN as \textit{Dn-RSN}.

%\RED{additionally investigate the effect }
% 
%learn the denoiser that directly maps the noisy image to the denoised image by removing the skip connection in SimpleCNN, and apply RSN in the learning process. We name this learned denoiser with RSN as \textit{Dn-RSN}.

In training the aforementioned NNs at each Momentum-Net layer, the parameters of the current layer NN are initialized with those trained in the previous layer, except for the first layer where we adopt the random initialization. For training loss function, we use the mean-square-error (MSE) between denoised images and reference images, i.e.,
$\theta^{(l+1)} = \argmin_{\theta} \bbE\| \x_{\text{ref}} - \cD_{\theta}(\x^{(l)}) \|_2^2$,
%\begin{equation}
%\theta^{(l+1)} = \argmin_{\theta} \bbE\| \x_{\text{ref}} - \cD_{\theta^{(l)}}(\x^{(l)}) \|_2^2,
%\vspace{-0.05in}
%\end{equation}
where $\x_{\text{ref}}$ is a high-quality reference image.

We apply the $\rho$-relaxation strategy to image refining modules \cite{chun2019momentumnet}:  
\begin{equation}\label{eq:z}
\z^{(l+1)} = (1-\rho)(\x^{(l)}) + \rho \cD_{\mb{\theta}^{(l+1)}}(\x^{(l)}), \quad \rho\in(0,1).
\end{equation}
The $\rho$-relaxation strategy mixes the information of a reconstructed and denoised image, improving the MBIR performance, when the measurement noise is moderate or the imaging system is moderately ill-posed. We observed that $\rho=0.5$ improves LDCT image reconstruction accuracy than $\rho=1-\epsilon$, where $\epsilon$ is a machine epsilon. This correspond to the sparse-view CT experiment results in \cite{chun2019momentumnet}.

\subsection{Extrapolation module}	
Many optimization literatures have shown that using the momentum of previous updates can accelerate the algorithm convergence by smoothing ``zig-zagging" trajectories \cite{Nesterov13, Chun&Fessler:18TIP}. 
In particular, the momentum terms, $\{ \x^{(l)} - \x^{(l-1)} \}$, can amplify the changes in subsequent Momentum-Net layers \cite{chun2019momentumnet}.
The extrapolation modules give extrapolated points using the momentum terms:
\begin{equation}\label{eq:extrapolation}
\acute{\x}^{(l+1)} = \x^{(l)} + \delta^2m^{(l)}(\x^{(l)} - \x^{(l-1)}),
%\vspace{-0.05in}
\end{equation}
with $\delta=1-\epsilon$ and $\{0 \leq m^{(l)}\leq 1: l \geq 1 \}$.
%We call $\acute{\x}^{(l+1)}$ as the extrapolated image at the $(l+1)$th Momentum-Net layer. 
We update the momentum coefficients as follows \cite{chun2019momentumnet}: 
\begin{equation}\label{eq:mcoeff}
%\vspace{-0.05in}
m^{(l)} = \frac{t^{(l-1)}-1}{t^{(l)}}, \quad t^{(l)} = \frac{1+\sqrt{1+4(t^{(l-1)})^2}}{2},
%\vspace{-0.1in}
\end{equation}
where $t^{(0)} = 1$. 
This choice gave significant acceleration in solving several block optimization problems; see \cite{Chun&Fessler:18TIP} and references therein.

%\RED{In early Momentum-Net layers,} the reconstructed images often have dramatic changes across layers, and the momentum term is not necessarily activated. While with more layers, the changes of images across layers become moderate, and amplifying the changes from previous layers by emphasizing more on the momentum term in \eqref{eq:extrapolation} can accelerate the convergence. 
%We therefore design $m$ as 
\subsection{MBIR module}
We apply the PWLS method \cite{Fessler:94TMI} to the MBIR modules of Momentum-Net. At the $l\rth$ Momentum-Net layer, we consider the following MBIR problem:
\begin{equation}\label{eq:P0}
%\vspace{-0.05in}
\argmin_{\x\geq 0}F(\x; \y, \z^{(l+1)})\triangleq\frac{1}{2}\|\y - \A\x\|_{\W}^2 + \frac{\beta}{2}\|\x - \z^{(l+1)}\|^2,
%\tag{P0}
\end{equation} 
where $\A\in\mathbb{R}^{N_d \times N_p}$ is the system matrix and $\W$ is a diagonal weighting matrix whose diagonal elements are $w_i = \frac{y_i^2}{y_i + \sigma^2}$ for $i = 1,\ldots,N_d$, with the variance of the electronic noise $\sigma^2$ \cite{pre-post-log,Chun&Talavage:13Fully3D}.
The regularization parameter $\beta$ balances the data-fidelity term (the first term in \eqref{eq:P0}) and the prior estimate from the image refining module (the second term in \eqref{eq:P0}). 

%\BLUE{When reconstructing images with severely corrupted (low-dose) measurements, emphasizing more on the prior estimate, i.e., applying a larger $\beta$ to the regularizer, can better suppress the noise and artifacts. On the contrary, with mildly corrupted measurements, one can rely more on the data-fidelity by using a relatively small $\beta$.}

The MBIR module of Momentum-Net solves a \emph{majorized} version of MBIR problem \R{eq:P0} at the extrapolated point $\acute{\x}^{(l+1)}$, and the reconstructed image $\x^{(l+1)}$ is obtained as
\begingroup
\setlength{\thinmuskip}{1.5mu}
\setlength{\medmuskip}{2mu plus 1mu minus 2mu}
\setlength{\thickmuskip}{2.5mu plus 2.5mu}
%\fontsize{9.5pt}{11.4pt}\selectfont
\allowdisplaybreaks
\eas{
&~ \mb{x}^{(l+1)}
\\
&= \argmin_{\mb{x} \geq 0} \frac{1}{2} \left\| \mb{x} - \left( \acute{\x}^{(l+1)} - \mb{M}^{-1}\triangledown F(\acute{\x}^{(l+1)}; \y, \z^{(l+1)}) \right) \right\|_{\mb{M}}^2
%\BLUE{\text{Prox}_{\bbI_{\cX}}^{\mb{M}}\big(\acute{\x}^{(l+1)} - \mb{M}^{-1}\triangledown F(\acute{\x}^{(l+1)}; \y, \z^{(l+1)})\big)}\\
\\
&= \big[\acute{\x}^{(l+1)} - \mb{M}^{-1}\triangledown F(\acute{\x}^{(l+1)}; \y, \z^{(l+1)})\big]_+,
}
\endgroup
where ${\mb{M} = \A^T\W\A\bm{1} + \beta\mb{I}}$ is the majorization matrix for $\nabla F(\mb{x}; \mb{y}, \mb{z}^{(l+1)})$ that is given in \R{eq:P0} \cite{Chun&Fessler:18TIP,Chun&Fessler:20TIP}, the operator $[\cdot]_{+}$ sets negative elements as zeros, and $\bm{1}$ is a vector with ones.

\section{Experimental Results}
\label{sec:experiments}
We trained three Momentum-Nets with different combinations of image refining NN architectures and learning methods, i.e., SimpleCNN \cite{ryu2019plug}, SimpleCNN-RSN, and Dn-RSN, as described in Sec.~\ref{sec:refining}. We compared the performance of the trained Momentum-Nets with a state-of-the-art LDCT denoising method, WavResNet \cite{WavResNet18}, and examined the effect of learning nonexpansive mappings with RSN \cite{ryu2019plug} on Momentum-Net image reconstruction accuracy. We evaluated the results in terms of visual image qualities and the RMSE (root mean-square-errors) metric\footnote{This paper uses the shifted Hounsfield unit (HU) for RMSE, where air is 0 HU, and water is 1000 HU.}.

\subsection{Experimental setup}
\nin \emph{A. Imaging system and data:}
We evaluated the proposed method with the NIH AAPM-Mayo Clinic Low Dose CT Grand Challenge data \cite{MayoData}. 
The data includes standard-dose scans of ten patient examinations. 
We used the standard-dose CT images with 3~mm slice thickness as reference images, 
and simulated the low-dose post-log sinogram $\y$ using the Poisson-Gaussian model \cite{Chun&Talavage:13Fully3D,spultra}:
\begingroup
\setlength{\thinmuskip}{1.5mu}
\setlength{\medmuskip}{2mu plus 1mu minus 2mu}
\setlength{\thickmuskip}{2.5mu plus 2.5mu}
%\fontsize{9.5pt}{11.4pt}\selectfont
\bes{
y_{i} = - \log \left( I_0^{-1}\max\big( \text{Poisson} (I_0 e^{-[\A\x_{\text{ref}}]_i}) + \mathcal{N}(0, \sigma^2),\epsilon \big) \right),
}
\endgroup
for $ i = 1,\ldots,N_d $, with the number of incident photons per ray, $I_0=1\times10^4$, and the variance of electronic noise, $\sigma^2=25$ \cite{pre-post-log,spultra}.
We generated the sinograms with 2D fan-beam CT geometry using 736 `detectors or rays'~$\times$~1152 `regularly space projection views or angles', and no-scatter monoenergetic source.
The reconstructed or denoised images are of size ${512\times 512}$ at a resolution of 0.69~mm $\times$ 0.69~mm. 

We used FBP images as (initial) inputs to Momentum-Nets and WavResNet. In Momentum-Net training, we used 120 training pairs from six out of ten patients: each pair consists of reconstructed low-dose images and standard-dose reference images. In training the noniterative deep NN, WavResNet, we used all the 1466 low-dose FBP and reference image pairs from the same six training patients, in order to reduce overfitting risks. We tested all these methods with another 12 low-dose scans collected from the remaining four patients.

\vspace{0.5pc}

\nin \emph{B. Momentum-Net parameters and training:}
We set $\rho = 0.5$ in the Momentum-Net extrapolation modules. In MBIR modules, we used the adaptive regularization parameter selection scheme based on the spectral radius \cite{chun2019momentumnet} to provide a distinct $\beta$ value for each sample. We chose $119$ for the \BLUE{desired factor $\chi^\star$}.
\begin{figure}[!t]
	\centering
	\begin{subfigure}{0.23\textwidth}
		\includegraphics[width=1\textwidth]{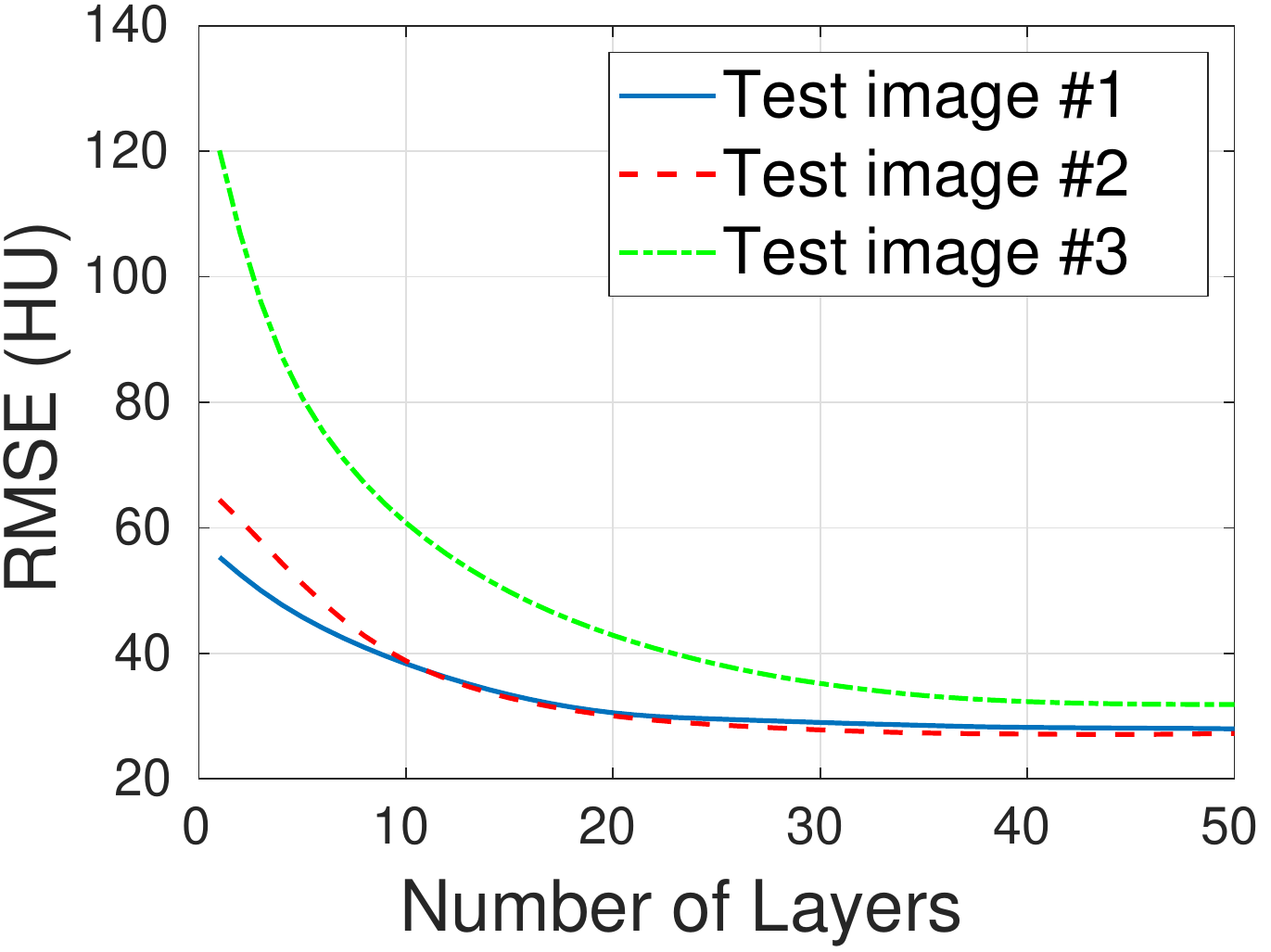}
%		\vspace{-1.5pc}
%		\caption{RMSE examples}
		\label{fig:RMSE-3pat}
	\end{subfigure}
	\hfil
	\begin{subfigure}{0.23\textwidth}		\includegraphics[width=1\textwidth]{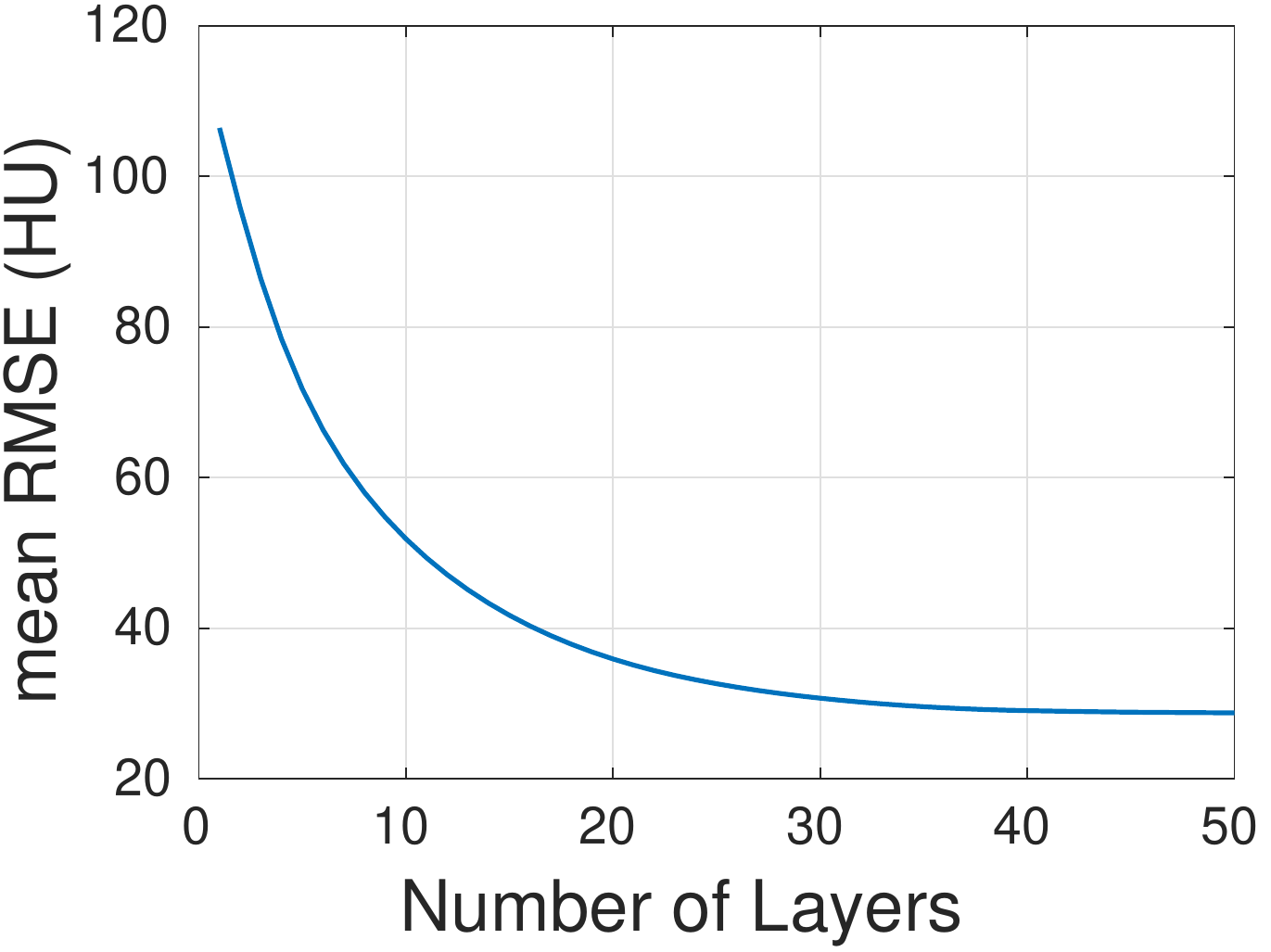}
%	\vspace{-1.5pc}
%	\caption{Mean RMSE}
		\label{fig:meanRMSE}
	\end{subfigure}
	\vspace{-1.5pc}
	\caption{RMSE minimization behavior of Momentum-Net with SimpleCNN: (Left) RMSEs of three examples, (Right) Mean RMSE of 12 test images.}
	\label{fig: num-3pat}
	\vspace{-0.5pc}
\end{figure}

\begin{table}[!t]
	\centering
	\caption{Mean and standard deviation (STD) of RMSE (HU) of 12 reconstructed test images with different methods.}
	\vspace{-0.8pc}
	\begin{tabular}{c|cc}
		\hline
		Method&\multicolumn{1}{c}{Mean}&\multicolumn{1}{c}{STD} \\ \hline \hline
		FBP &117.85 &58.891   \\ \hline
		WavResNet        &33.25 & 6.46      \\  \hline
		\begin{tabular}[c]{@{}c@{}}\bfseries Momentum-Net (SimpleCNN)\end{tabular}  & {\bfseries 28.77} & {\bfseries 3.55}  \\  \hline
		\begin{tabular}[c]{@{}c@{}}Momentum-Net (SimpleCNN-RSN)\end{tabular} & 36.10 & 4.87  \\  \hline
		\begin{tabular}[c]{@{}c@{}}Momentum-Net (Dn-RSN)\end{tabular} & 41.72 &7.14   \\  \hline
	\end{tabular}
\vspace{-0.5pc}
	\label{tab:numerical}
	\vspace{-0.5pc}
\end{table}
In training the three Momentum-Nets, we decreased the learning rate from $10^{-3}$ by a ratio of $0.9$ every ten epochs, and used Adam \cite{Kingma&Ba:15ICLR} to optimize network parameters. We performed image-based training with the mini-batch size of 5 and 100 epochs in each Momentum-Net layer. The three Momentum-Nets used the same dimensions of convolutional kernels as \cite{ryu2019plug}, i.e., the first layer uses 64 filters of size ${3\times 3}$, the second and third layer uses 64 filters of size ${3 \times 3 \times 64}$, and the last layer uses one filter of size ${3\times 3\times 64}$ to reconstruct the output. 

\begin{figure*}[!t]
	\centering	
	\begin{subfigure}{1\textwidth}
\scalebox{0.82}{\begin{tikzpicture}
	[spy using outlines={rectangle,green,magnification=2.3,size=10mm, connect spies}]
	\node {\includegraphics[width=0.22\textwidth]{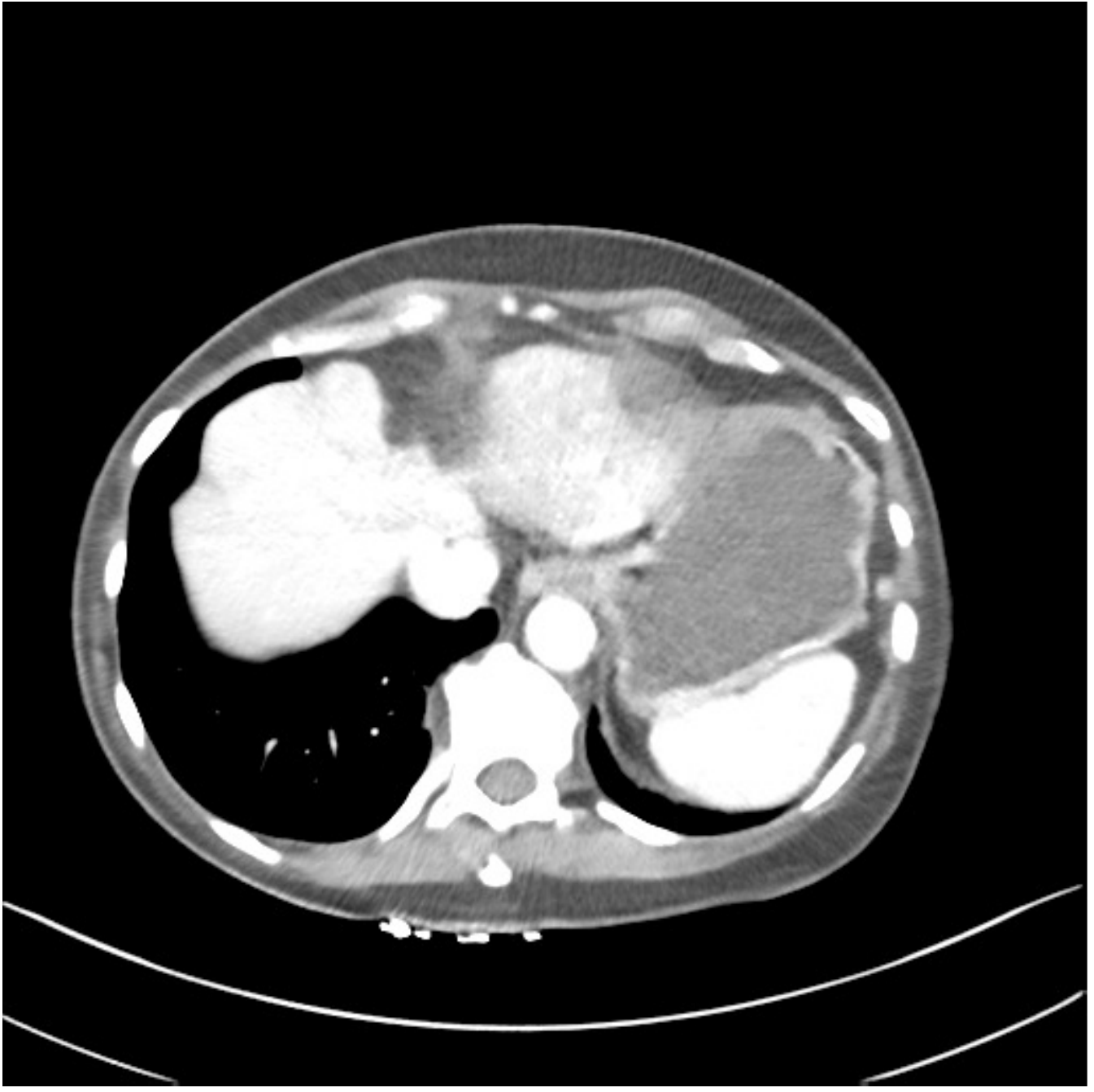} };
	\spy on (0.2,-0.1) in node [right] at (0.9,1.4);
	\spy on (-0.40,-1.25) in node [left] at (-1,-1.65);
	\node [white, font=\bf] at (-0.5,1.6) {WavResNet \cite{WavResNet18}};
	\end{tikzpicture}
	\begin{tikzpicture}
	[spy using outlines={rectangle,green,magnification=2.3,size=10mm, connect spies}]
	\node {\includegraphics[width=0.22\textwidth]{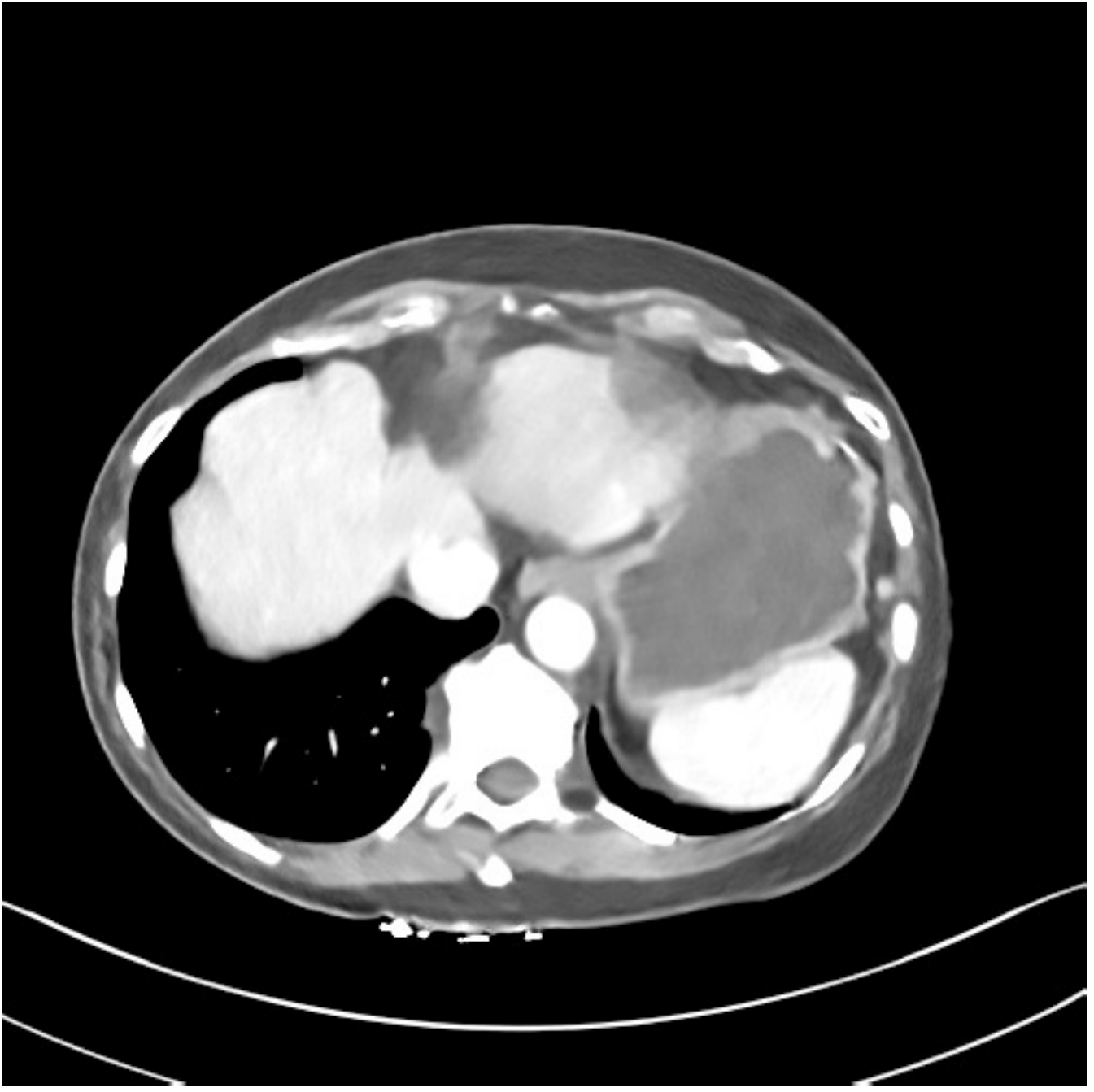} };
		\spy on (0.2,-0.1) in node [right] at (0.9,1.4);
	\spy on (-0.40,-1.25) in node [left] at (-1,-1.65);
	\node [align = center,white, font=\bf] at (-0.5,1.52) {\small Momentum-Net \\ (SimpleCNN)};
	\end{tikzpicture}
	\begin{tikzpicture}
	[spy using outlines={rectangle,green,magnification=2.3,size=10mm, connect spies}]
	\node {\includegraphics[width=0.22\textwidth]{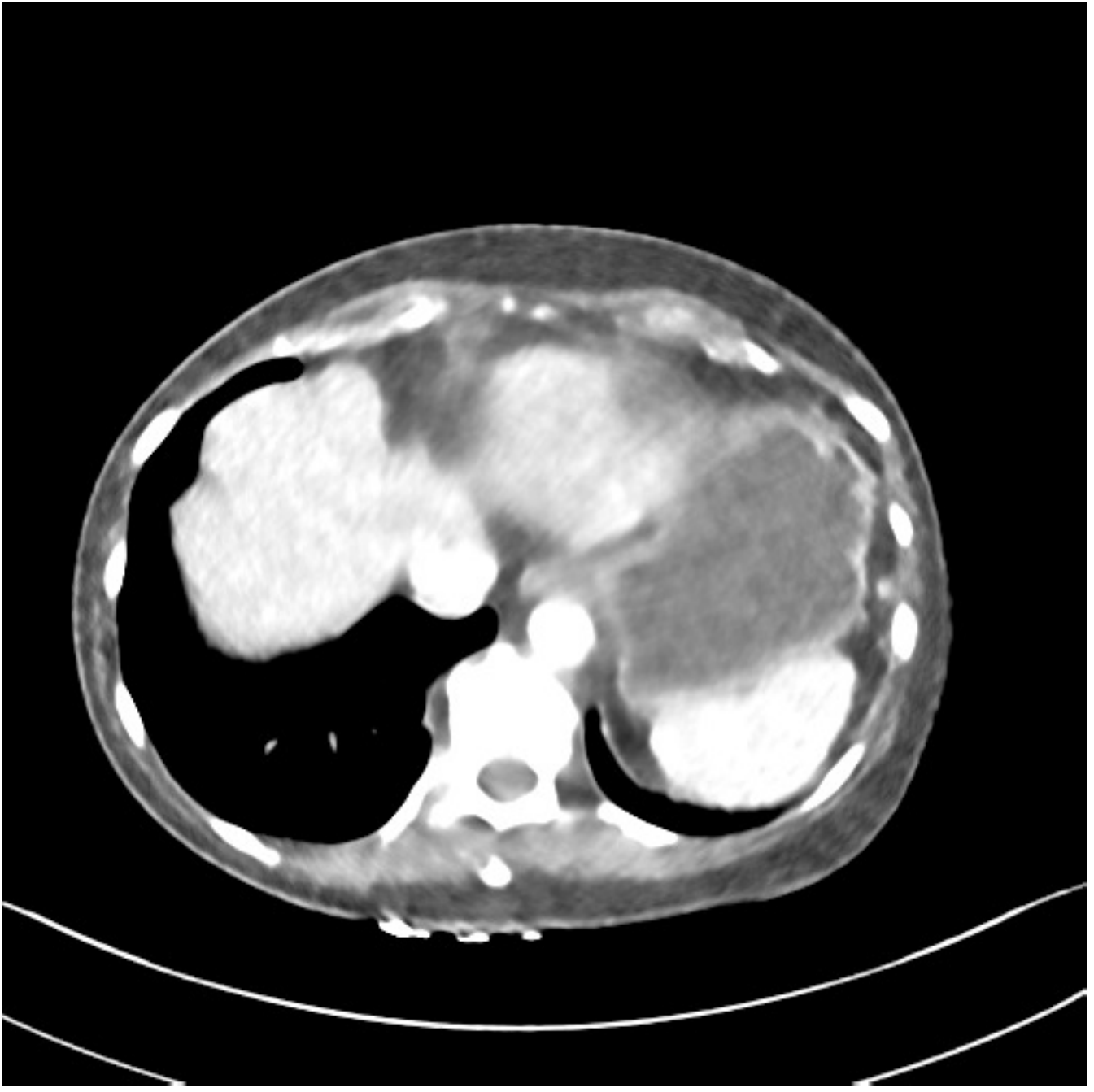} };
	\spy on (0.2,-0.1) in node [right] at (0.9,1.4);
	\spy on (-0.40,-1.25) in node [left] at (-1,-1.65);
	\node [align = center,white, font=\bf] at (-0.5,1.52) {\small Momentum-Net \\ (SimpleCNN-RSN)};
	\end{tikzpicture}
	\begin{tikzpicture}
	[spy using outlines={rectangle,green,magnification=2.3,size=10mm, connect spies}]
	\node {\includegraphics[width=0.22\textwidth]{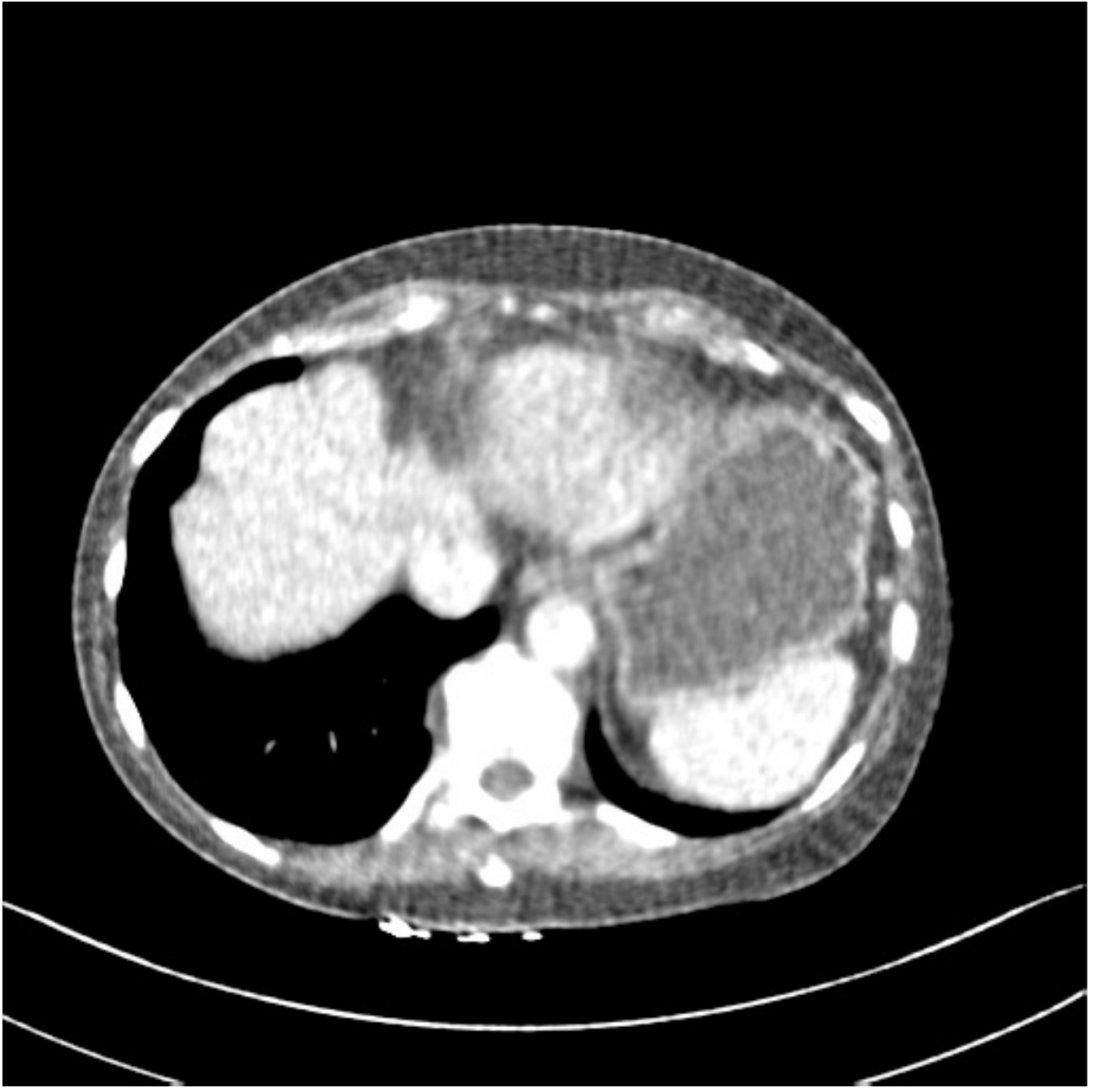} };
	\spy on (0.2,-0.1) in node [right] at (0.9,1.4);
	\spy on (-0.40,-1.25) in node [left] at (-1,-1.65);
	\node [align = center,white, font=\bf] at (-0.5,1.52) {\small Momentum-Net \\ (Dn-RSN)};
	\end{tikzpicture}
	\begin{tikzpicture}
	[spy using outlines={rectangle,green,magnification=2.3,size=10mm, connect spies}]
	\node {\includegraphics[width=0.22\textwidth]{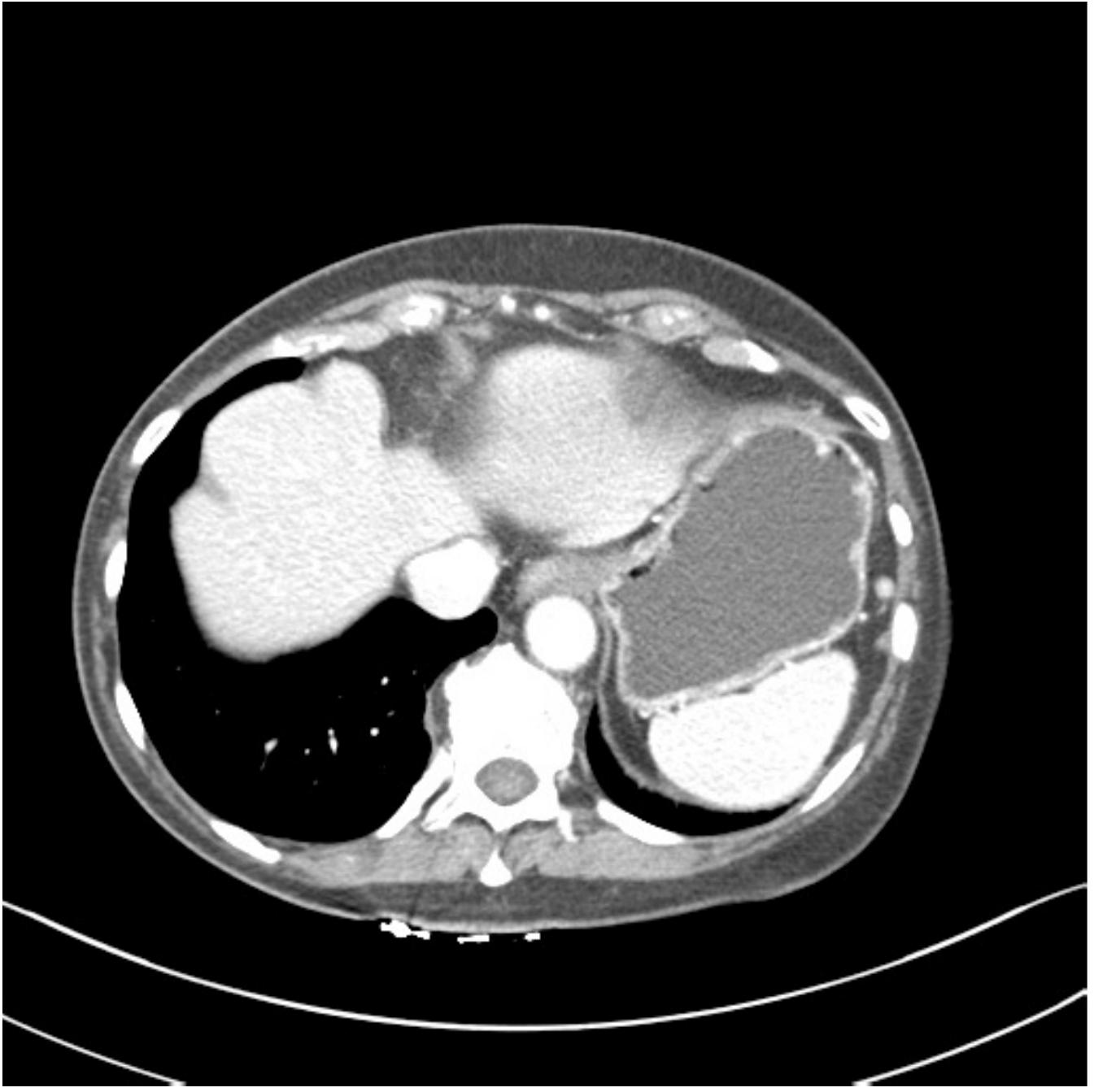} };
		\spy on (0.2,-0.1) in node [right] at (0.9,1.4);
	\spy on (-0.40,-1.25) in node [left] at (-1,-1.65);
	\node [white, font=\bf] at (-0.3,1.6) {Reference};
	\end{tikzpicture}}
%	\caption{Testing image \# 1}
	\label{fig:pat1-1}	
\end{subfigure}
\vfil \vspace{-0.2pc}
\begin{subfigure}{1\textwidth}
\scalebox{0.82}{\begin{tikzpicture}\begin{scope}
	[spy using outlines={rectangle,green,magnification=2.3,size=10mm, connect spies}]
	\node {\includegraphics[width=0.22\textwidth]{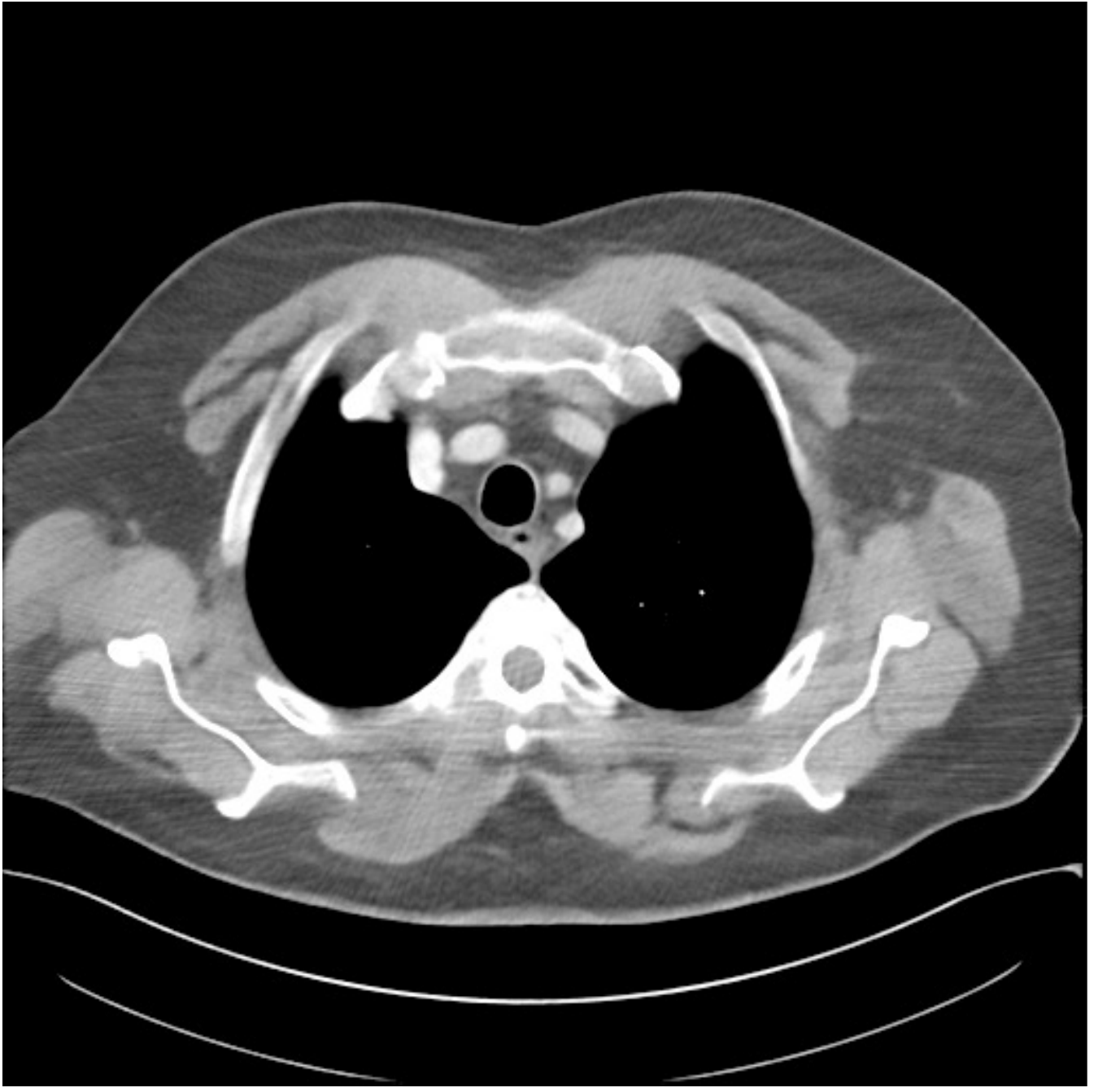} };
	\spy on (1.5,0.05) in node [right] at (0.9,1.4);
	\spy on (-1.5,-0.4) in node [left] at (-1,-1.5);
		\end{scope}
	\draw [->,>=stealth,red,line width=0.8pt] (-1.4,-1.9) -- (-1.4,-1.55);
\draw [->,>=stealth,red,line width=0.8pt] (1.8,1.8) -- (1.45,1.7);
\draw [->,>=stealth,red,line width=0.8pt] (1.7,1.4) -- (1.35,1.4);
%	\node [white, font=\bf] at (-0.5,1.6) {WavResNet \cite{WavResNet18}};
	\end{tikzpicture}
	\begin{tikzpicture}\begin{scope}
	[spy using outlines={rectangle,green,magnification=2.3,size=10mm, connect spies}]
	\node {\includegraphics[width=0.22\textwidth]{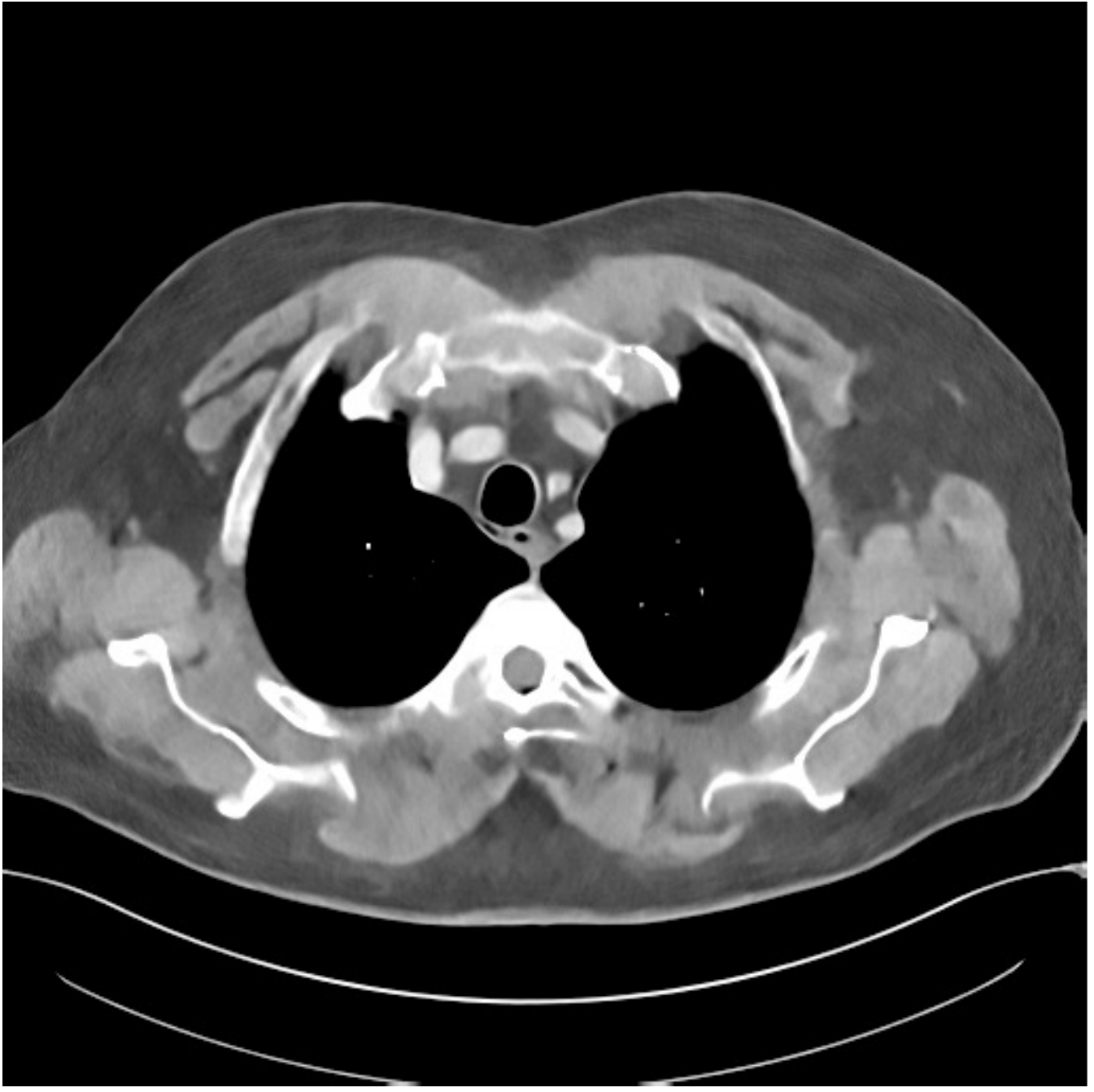} };
	\spy on (1.5,0.05) in node [right] at (0.9,1.4);
	\spy on (-1.5,-0.4) in node [left] at (-1,-1.5);
		\end{scope}
	\draw [->,>=stealth,red,line width=0.8pt] (-1.4,-1.9) -- (-1.4,-1.55);
    \draw [->,>=stealth,red,line width=0.8pt] (1.8,1.8) -- (1.45,1.7);
    \draw [->,>=stealth,red,line width=0.8pt] (1.7,1.4) -- (1.35,1.4);
%	\node [align=center,white, font=\bf] at (-0.5,1.6) {Momentum-Net \\ (SimpleCNN)};
	\end{tikzpicture}
	\begin{tikzpicture}\begin{scope}
	[spy using outlines={rectangle,green,magnification=2.3,size=10mm, connect spies}]
	\node {\includegraphics[width=0.22\textwidth]{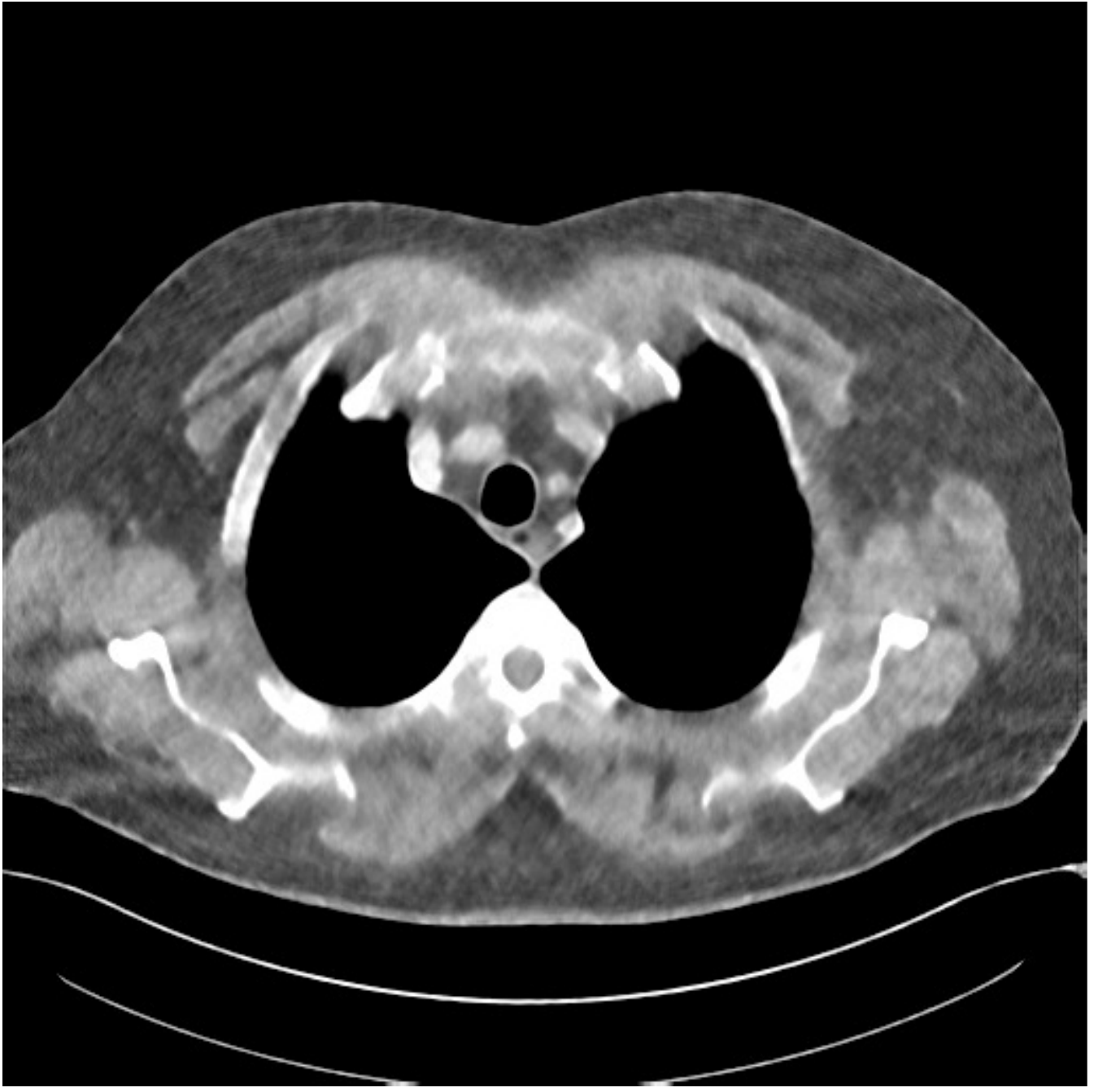} };
	\spy on (1.5,0.05) in node [right] at (0.9,1.4);
\spy on (-1.5,-0.4) in node [left] at (-1,-1.5);
	\end{scope}
\draw [->,>=stealth,red,line width=0.8pt] (-1.4,-1.9) -- (-1.4,-1.55);
\draw [->,>=stealth,red,line width=0.8pt] (1.8,1.8) -- (1.45,1.7);
\draw [->,>=stealth,red,line width=0.8pt] (1.7,1.4) -- (1.35,1.4);
%	\node [align = cenred,ite, font=\bf] at (-0.5,1.52) {\small Momentum-Net \\ (SimpleCNN-RSN)};
	\end{tikzpicture}
	\begin{tikzpicture}\begin{scope}
[spy using outlines={rectangle,green,magnification=2.3,size=10mm, connect spies}]
\node {\includegraphics[width=0.22\textwidth]{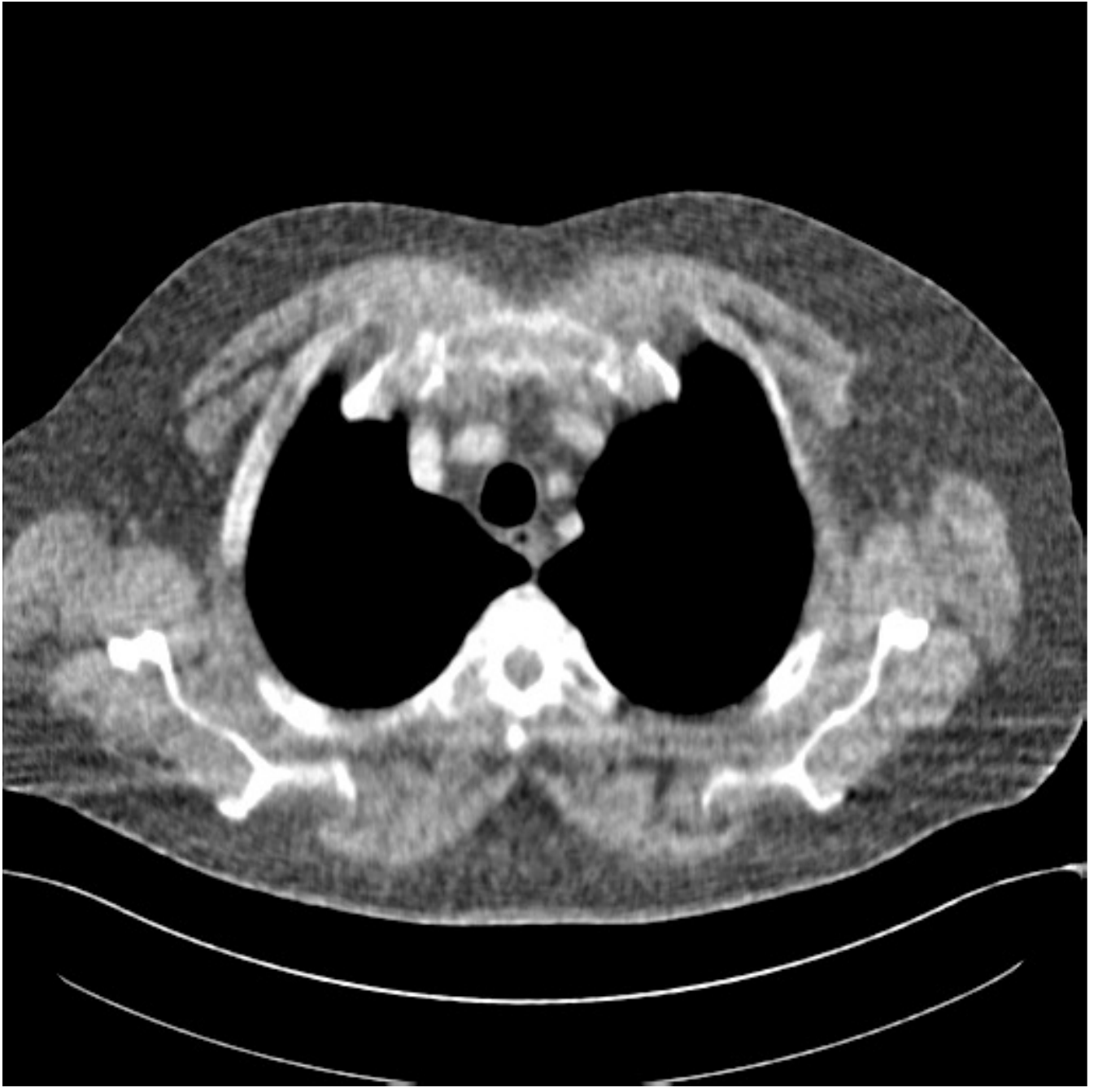} };
	\spy on (1.5,0.05) in node [right] at (0.9,1.4);
\spy on (-1.5,-0.4) in node [left] at (-1,-1.5);
	\end{scope}
	\draw [->,>=stealth,red,line width=0.8pt] (-1.4,-1.9) -- (-1.4,-1.55);
	\draw [->,>=stealth,red,line width=0.8pt] (1.8,1.8) -- (1.45,1.7);
	\draw [->,>=stealth,red,line width=0.8pt] (1.7,1.4) -- (1.35,1.4);
%\node [align = center,white, font=\bf] at (-0.5,1.52) {\small Momentum-Net \\ (Dn-RSN)};
\end{tikzpicture}
	\begin{tikzpicture}\begin{scope}
	[spy using outlines={rectangle,green,magnification=2.3,size=10mm, connect spies}]
	\node {\includegraphics[width=0.22\textwidth]{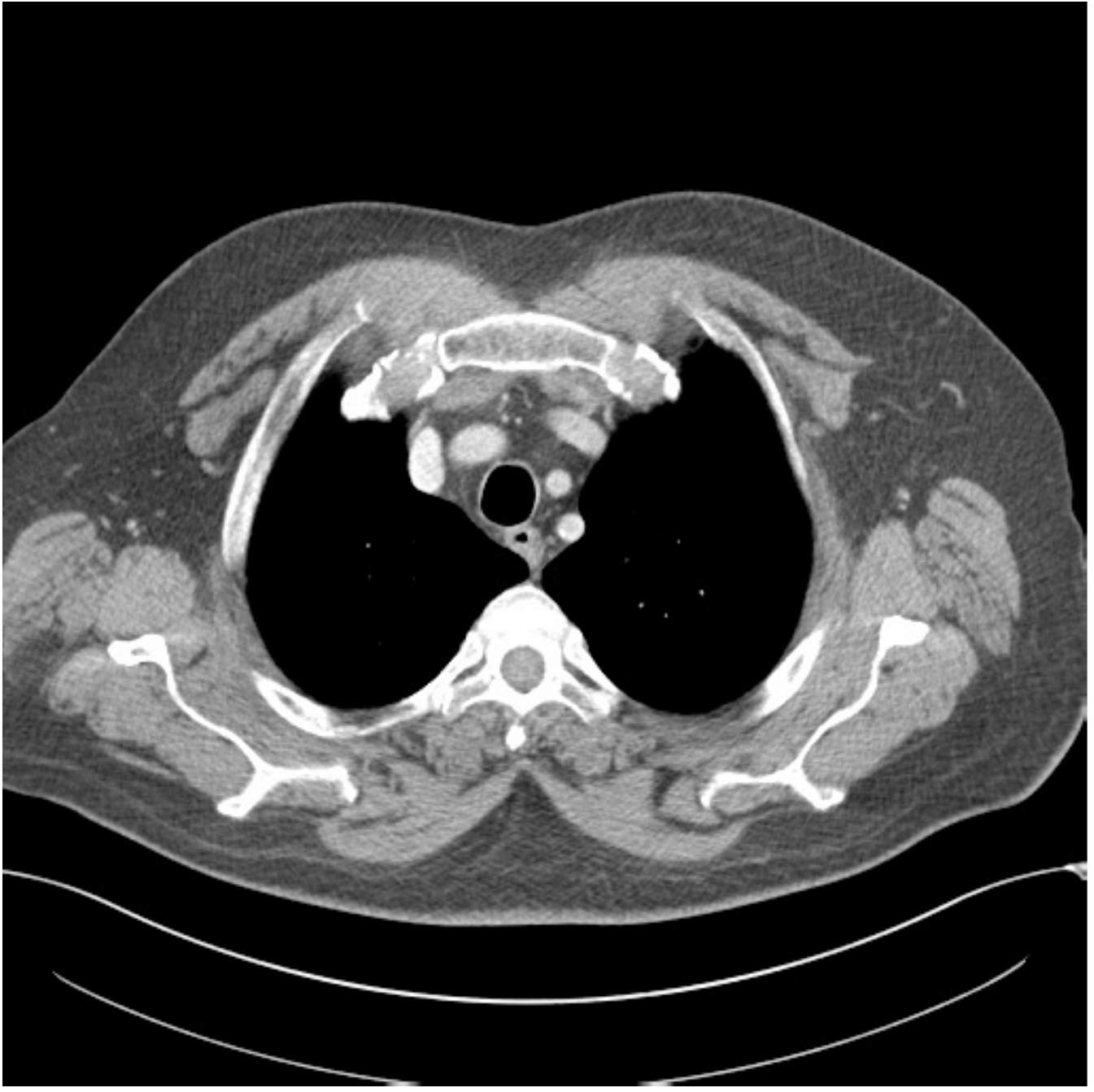} };
	\spy on (1.5,0.05) in node [right] at (0.9,1.4);
	\spy on (-1.5,-0.4) in node [left] at (-1,-1.5);
	\end{scope}
	\draw [->,>=stealth,red,line width=0.8pt] (-1.4,-1.9) -- (-1.4,-1.55);
	\draw [->,>=stealth,red,line width=0.8pt] (1.8,1.8) -- (1.45,1.7);
	\draw [->,>=stealth,red,line width=0.8pt] (1.7,1.4) -- (1.35,1.4);
%	\node [white, font=\bf] at (-0.3,1.6) {Reference};
	\end{tikzpicture}}
%\caption{Testing image \# 2}
	\label{fig:pat2-1}	
\end{subfigure}
\vfill \vspace{-0.3pc}
\begin{subfigure}{1\textwidth}
\scalebox{0.82}{\begin{tikzpicture}
	\begin{scope}
	[spy using outlines={rectangle,green,magnification=2.3,size=10mm, connect spies}]
	\node {\includegraphics[width=0.22\textwidth]{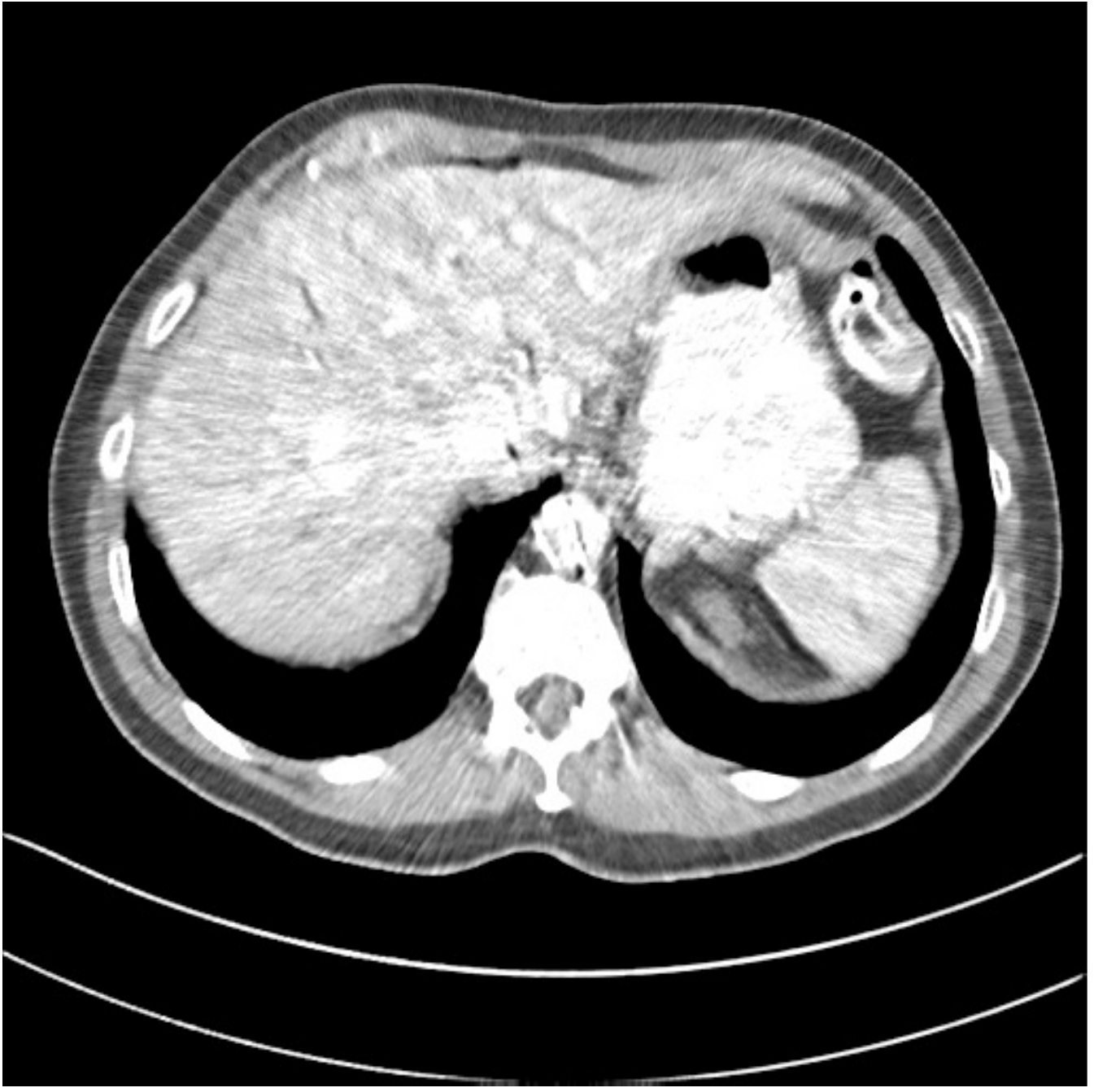} };
		\spy on (1.7,-0.05) in node [right] at (0.9,1.4);
	\spy on (-0.35,0.85) in node [left] at (-1,-1.5);
		\end{scope}
	\draw [->,>=stealth,red,line width=0.6pt] (-1.5,-1.6) -- (-1.55,-1.3);
	\draw [->,>=stealth,red,line width=0.6pt] (-1.6,-1.9) -- (-1.8,-1.7);
	\draw [->,>=stealth,red,line width=0.6pt] (-1.45,-1.85) -- (-1.15,-1.65);
%	\node [white, font=\bf] at (-0.5,1.8) {WavResNet \cite{WavResNet18}};
	\end{tikzpicture}
	\begin{tikzpicture}
	\begin{scope}
	[spy using outlines={rectangle,green,magnification=2.3,size=10mm, connect spies}]
	\node {\includegraphics[width=0.22\textwidth]{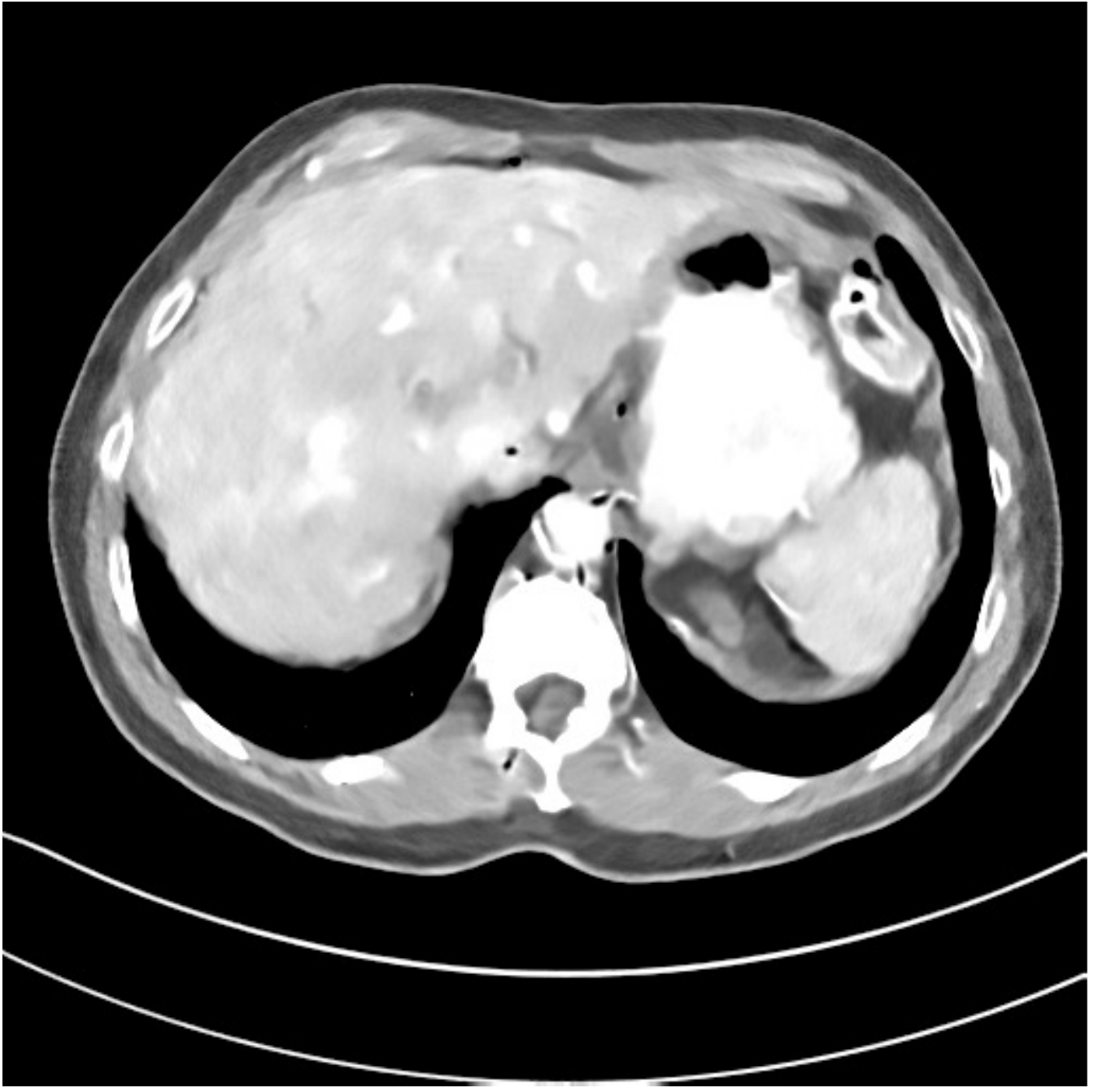} };
	\spy on (1.7,-0.05) in node [right] at (0.9,1.4);
	\spy on (-0.35,0.85) in node [left] at (-1,-1.5);
	\end{scope}
	\draw [->,>=stealth,red,line width=0.6pt] (-1.5,-1.6) -- (-1.55,-1.3);
	\draw [->,>=stealth,red,line width=0.6pt] (-1.6,-1.9) -- (-1.8,-1.7);
	\draw [->,>=stealth,red,line width=0.6pt] (-1.45,-1.85) -- (-1.15,-1.65);
%	\node [white, font=\bf] at (-0.5,1.8) {Momentum-Net};
	\end{tikzpicture}
		\begin{tikzpicture}
	\begin{scope}
	[spy using outlines={rectangle,green,magnification=2.3,size=10mm, connect spies}]
	\node {\includegraphics[width=0.22\textwidth]{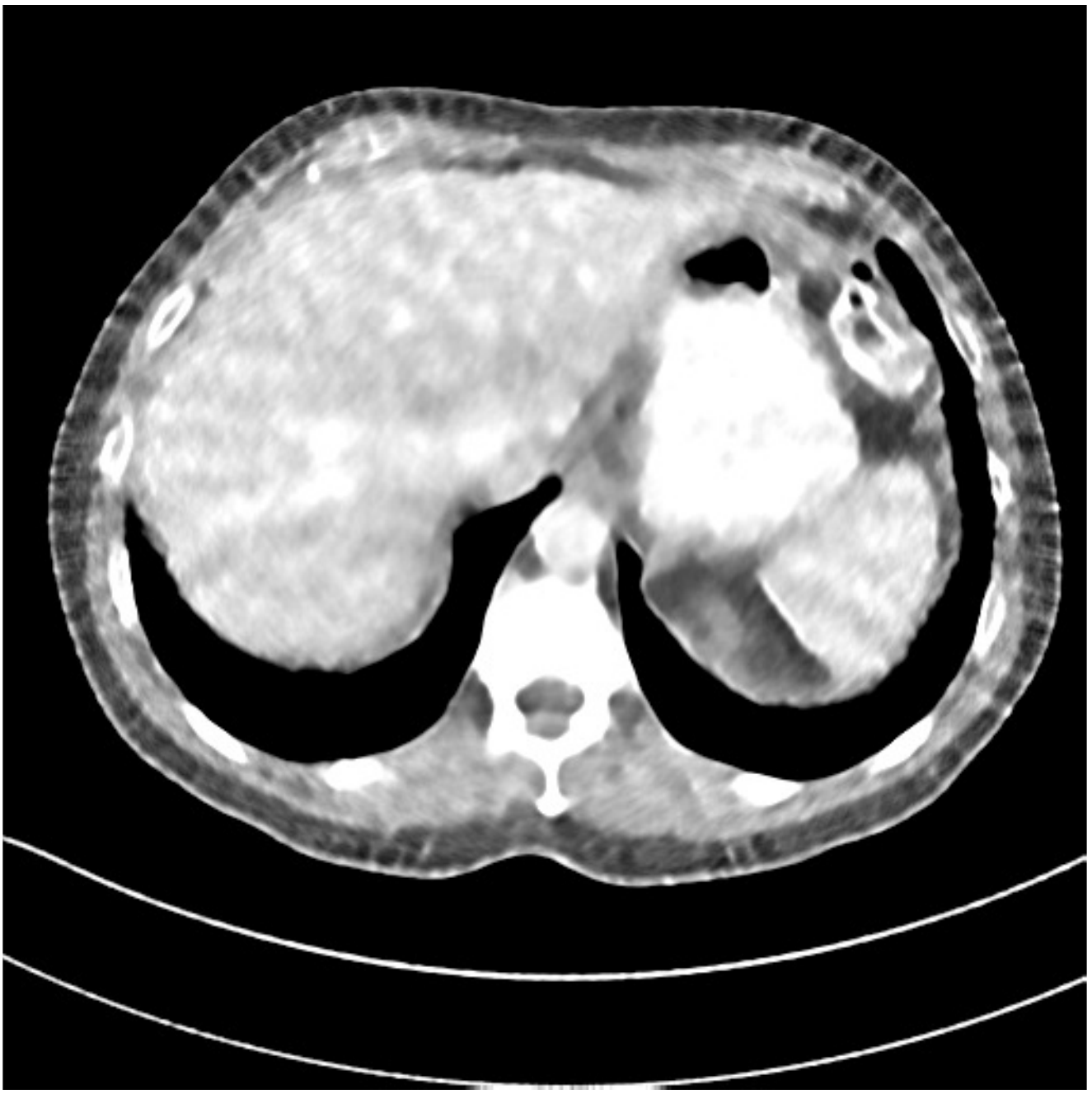} };
	\spy on (1.7,-0.05) in node [right] at (0.9,1.4);
	\spy on (-0.35,0.85) in node [left] at (-1,-1.5);
	\end{scope}
	\draw [->,>=stealth,red,line width=0.6pt] (-1.5,-1.6) -- (-1.55,-1.3);
	\draw [->,>=stealth,red,line width=0.6pt] (-1.6,-1.9) -- (-1.8,-1.7);
	\draw [->,>=stealth,red,line width=0.6pt] (-1.45,-1.85) -- (-1.15,-1.65);
	%	\node [white, font=\bf] at (-0.5,1.8) {Momentum-Net};
	\end{tikzpicture}
		\begin{tikzpicture}
   \begin{scope}
	[spy using outlines={rectangle,green,magnification=2.3,size=10mm, connect spies}]
	\node {\includegraphics[width=0.22\textwidth]{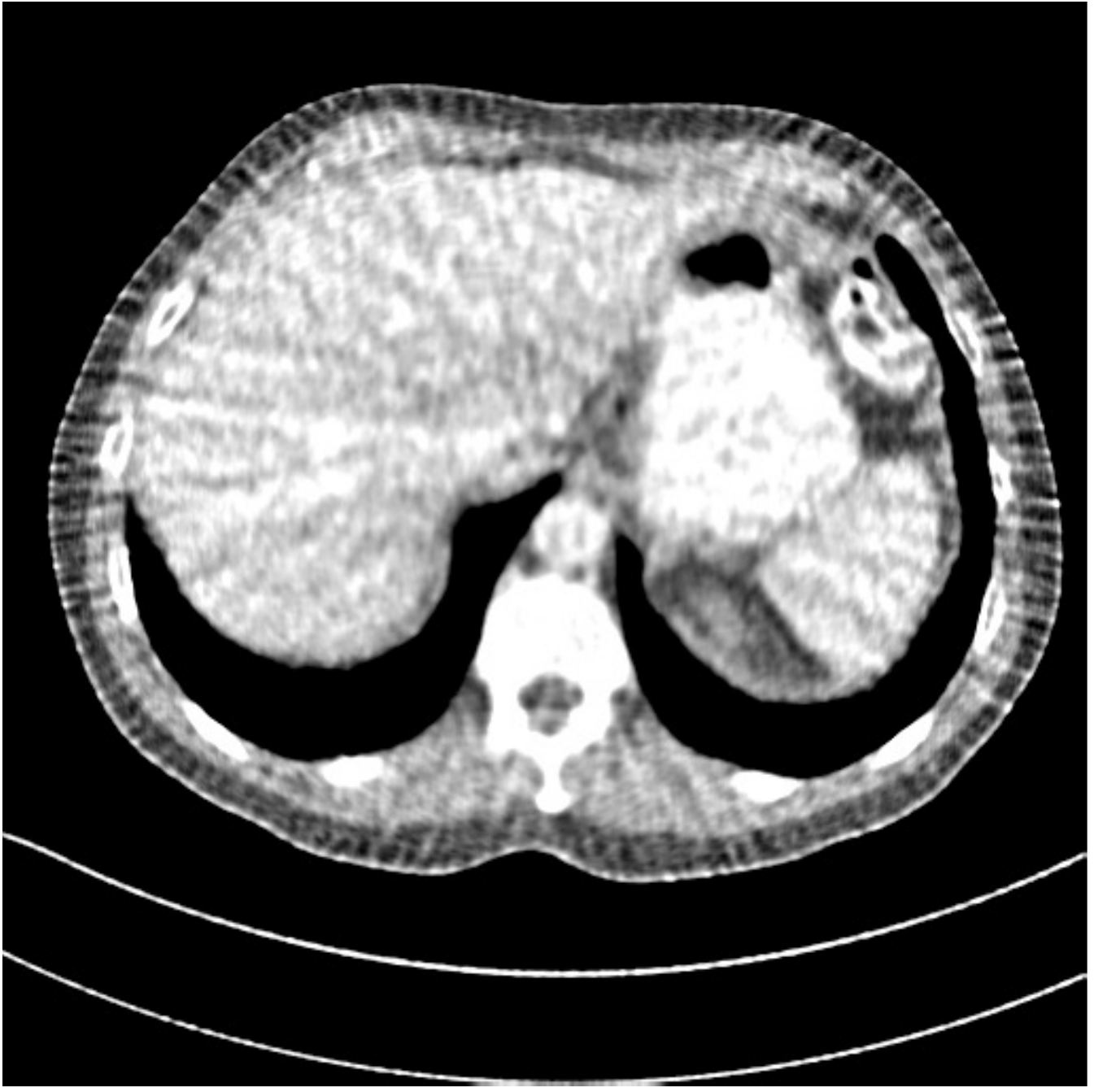} };
	\spy on (1.7,-0.05) in node [right] at (0.9,1.4);
	\spy on (-0.35,0.85) in node [left] at (-1,-1.5);
	\end{scope}
	 \draw [->,>=stealth,red,line width=0.6pt] (-1.5,-1.6) -- (-1.55,-1.3);
	\draw [->,>=stealth,red,line width=0.6pt] (-1.6,-1.9) -- (-1.8,-1.7);
	\draw [->,>=stealth,red,line width=0.6pt] (-1.45,-1.85) -- (-1.15,-1.65);
	%	\node [white, font=\bf] at (-0.5,1.8) {Momentum-Net};
	\end{tikzpicture}
	\begin{tikzpicture}
	\begin{scope}
	[spy using outlines={rectangle,green,magnification=2.3,size=10mm, connect spies}]
	\node {\includegraphics[width=0.22\textwidth]{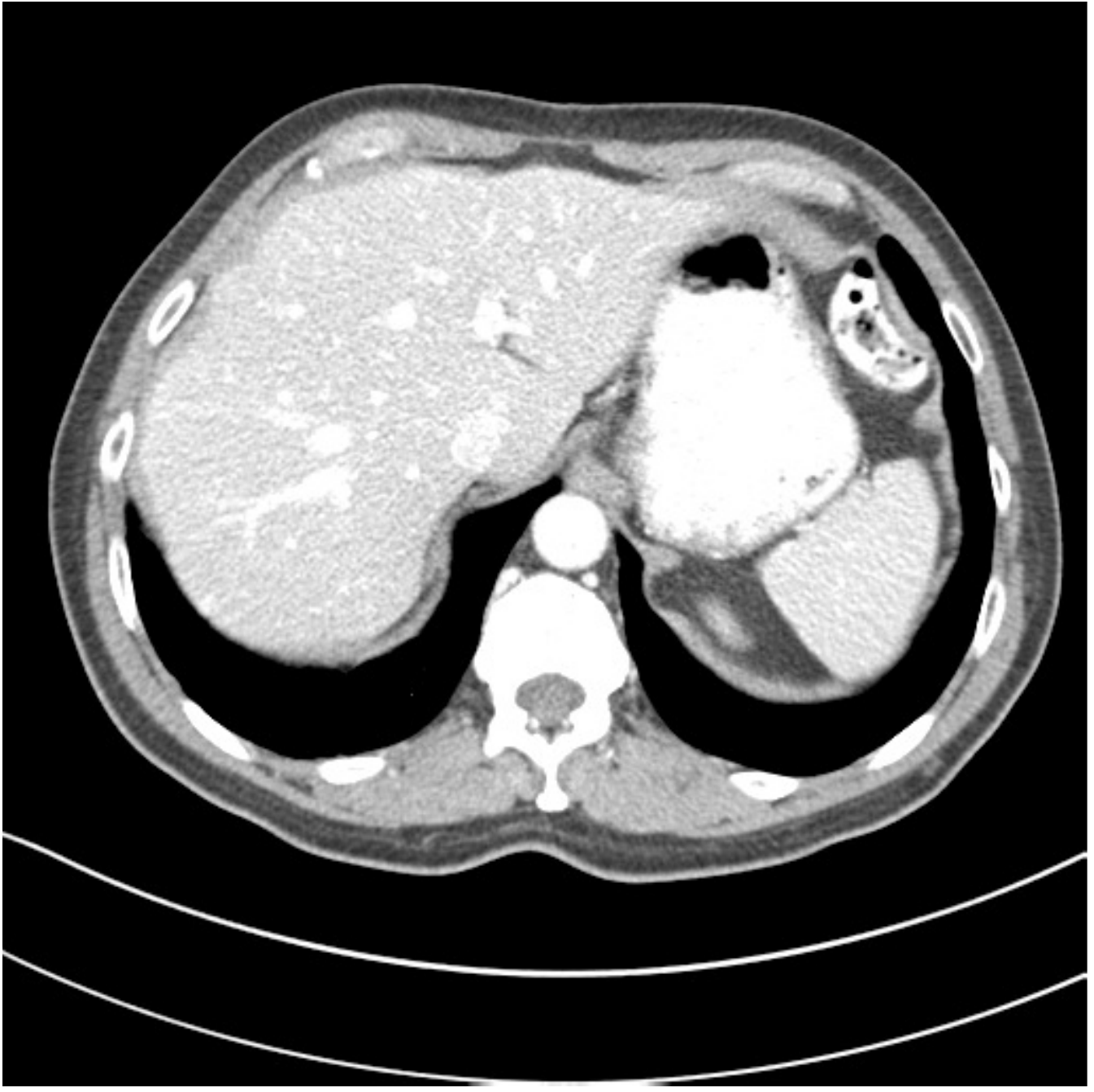} };
	\spy on (1.7,-0.05) in node [right] at (0.9,1.4);
	\spy  on (-0.35,0.85) in node [left] at (-1,-1.5);
	\end{scope}
	 \draw [->,>=stealth,red,line width=0.6pt] (-1.5,-1.6) -- (-1.55,-1.3);
	  \draw [->,>=stealth,red,line width=0.6pt] (-1.6,-1.9) -- (-1.8,-1.7);
	   \draw [->,>=stealth,red,line width=0.6pt] (-1.45,-1.85) -- (-1.15,-1.65);
	%	  \end{pgfonlayer}
	%	\node [white, font=\bf] at (-0.3,1.8) {Reference};
	\end{tikzpicture}
%	\begin{tikzpicture}
%	[spy using outlines={rectangle,green,magnification=2.3,size=10mm, connect spies}]
%	\node {\includegraphics[width=0.22\textwidth]{fig/mom_net_SimpleCNN/img/xtrue_test/pat3-1.pdf} };
%	\spy on (1.7,-0.05) in node [right] at (0.9,1.4);
%	\spy  on (-0.35,0.85) in node [left] at (-1,-1.5);
%%	 \begin{pgfonlayer}{foreground}
%%	 \draw [->,draw=red] (-2,-1) -- (-1,0);
%%	  \end{pgfonlayer}
%%	\node [white, font=\bf] at (-0.3,1.8) {Reference};
%	\end{tikzpicture}
   }
%\caption{Testing image \# 3}
	\label{fig:pat3-1}	
\end{subfigure}
\vspace{-0.6pc}
\caption{Three examples (from top to bottom) of the reconstructed testing images using Momentum-Net with SimpleCNN (the second column), with SimpleCNN-RSN (the third column), and with Dn-RSN (the fourth column). The compared WavResNet denoised images are shown in the first column, and the reference images are in the fifth column. See their FBP images in Fig.~\ref{fig:fbp}.}
\label{fig:mom-net-simplecnn}
\vspace{-0.7pc}
\end{figure*}
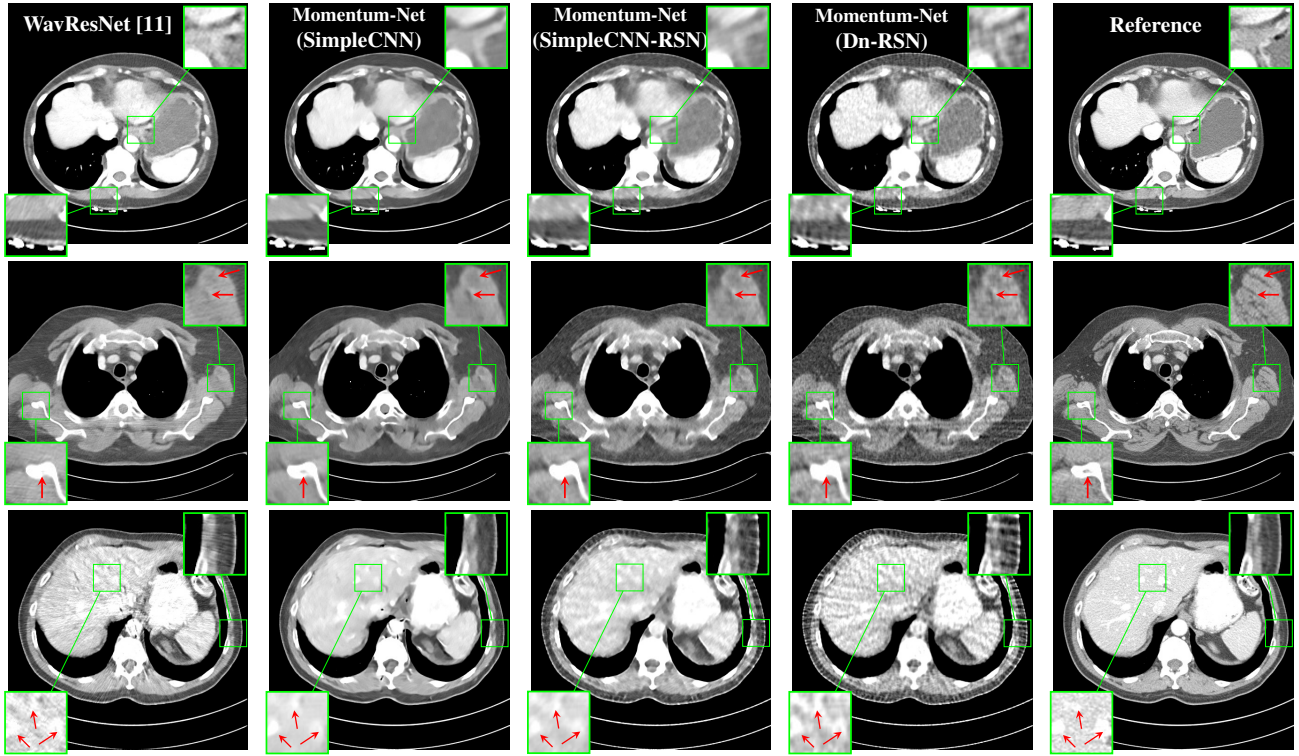

\begin{figure}[!h]
	\centering
	\begin{subfigure}{0.15\textwidth}
		\includegraphics[width=1\textwidth]{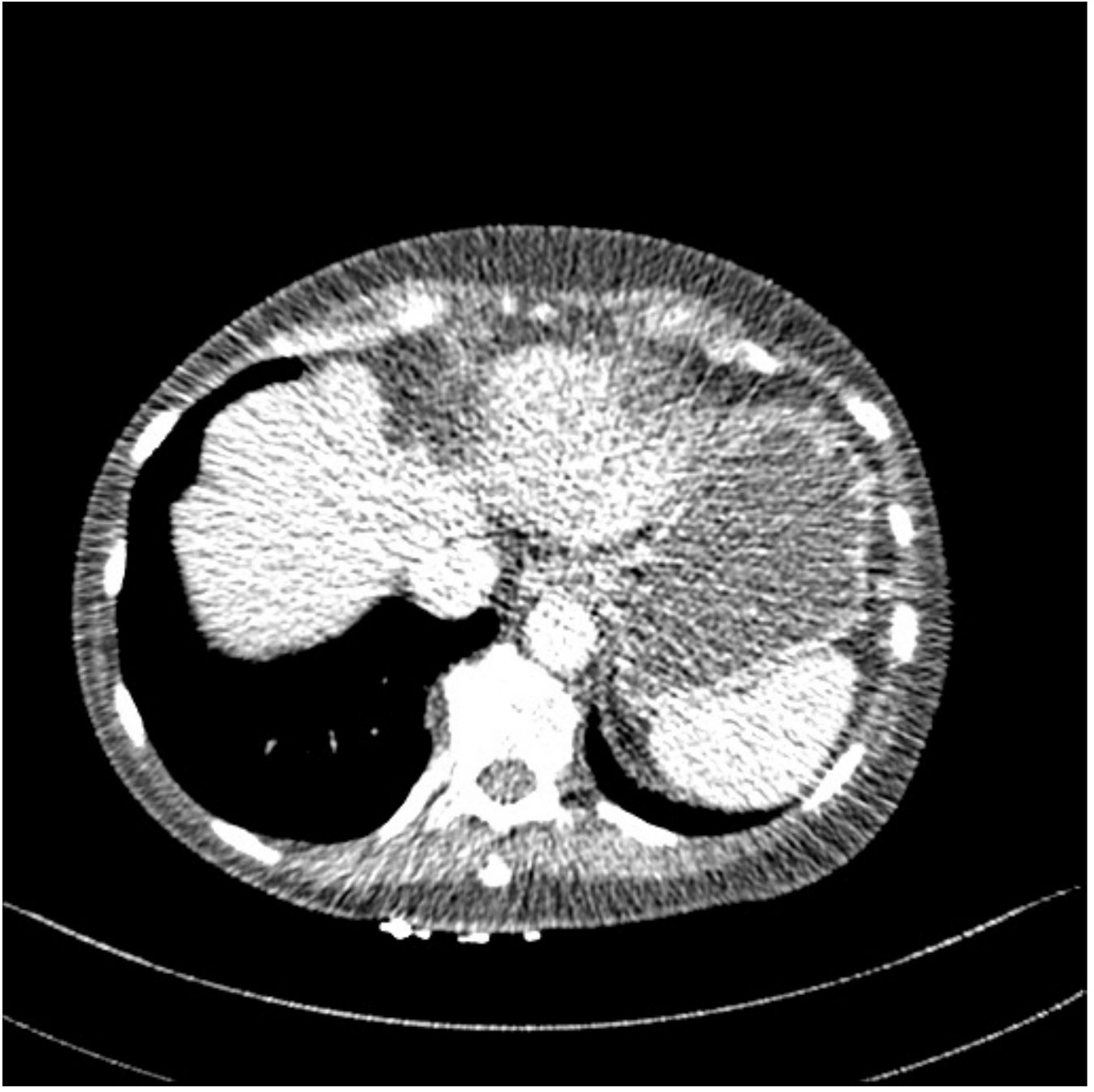}
		\label{fig:fbp-1}
	\end{subfigure}
	\hfil
	\begin{subfigure}{0.15\textwidth}
		\includegraphics[width=1\textwidth]{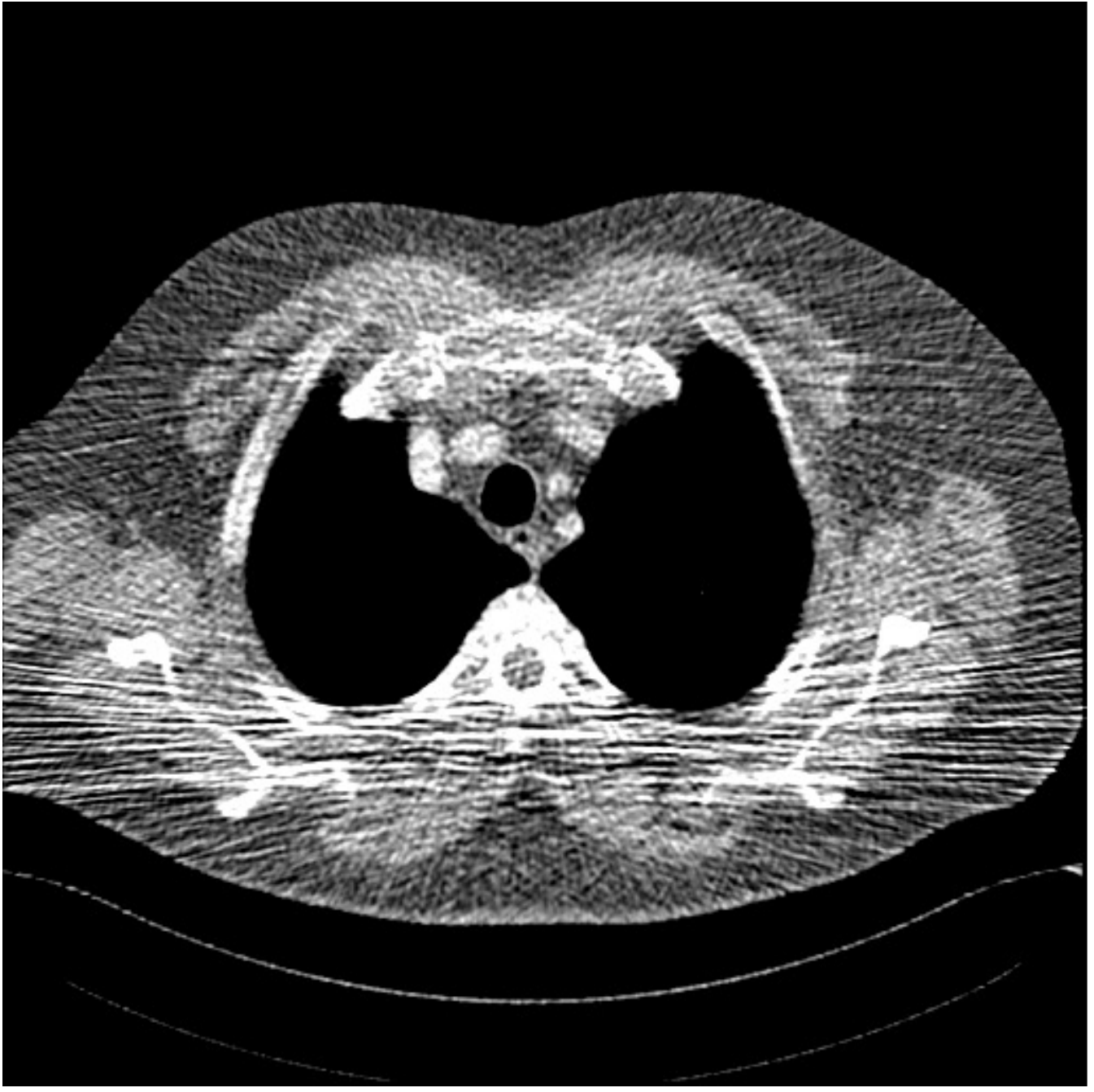}
		\label{fig:fbp-2}
	\end{subfigure}
	\hfil
	\begin{subfigure}{0.15\textwidth}
		\includegraphics[width=1\textwidth]{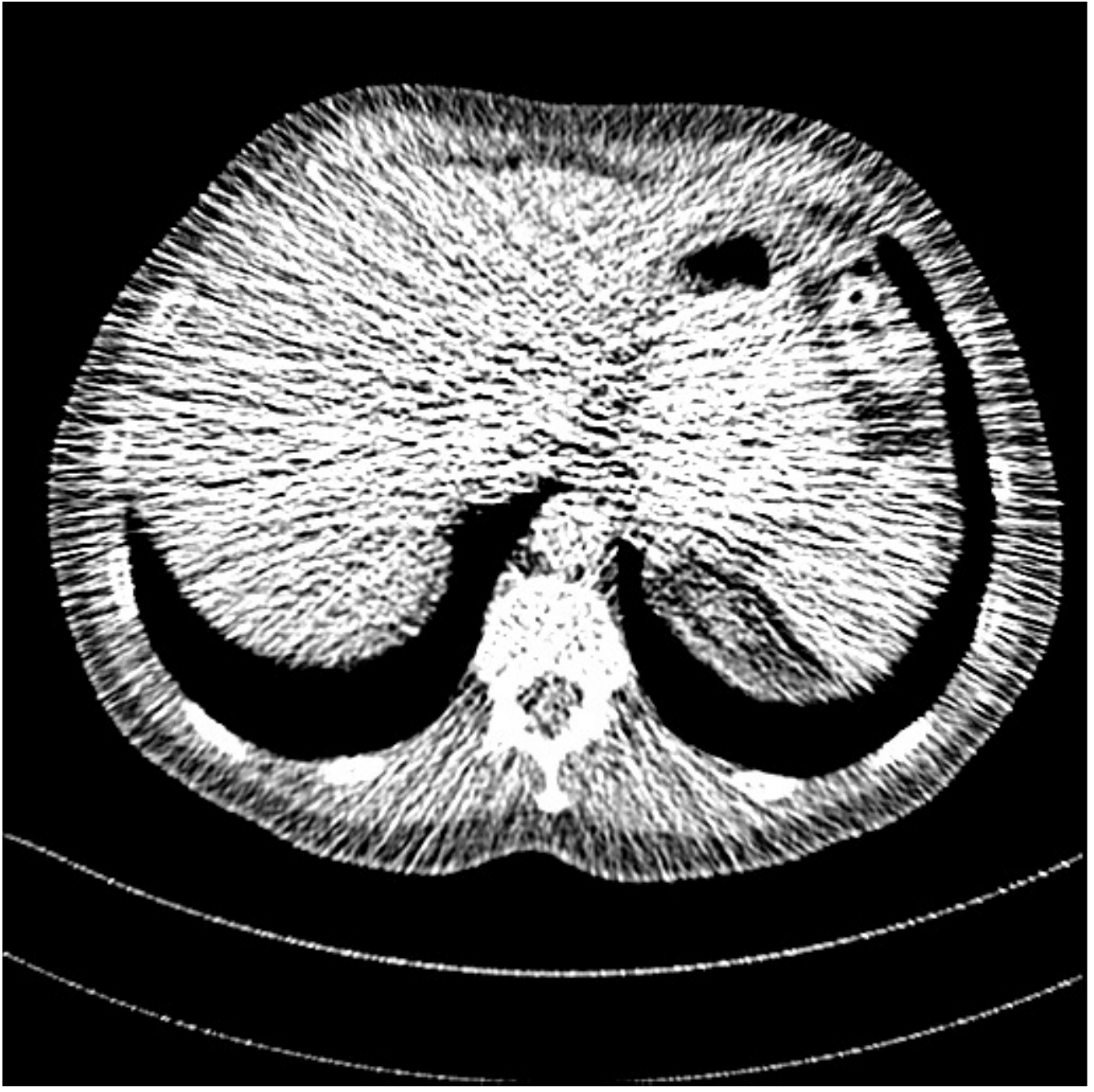}
		\label{fig:fbp-3}
	\end{subfigure}
	\vspace{-1.5pc}
	\caption{FBP images of test examples.}
	\label{fig:fbp}
	\vspace{-1.5pc}
\end{figure}
\subsection{Proposed Momentum-Net with SimpleCNN}
Fig.~\ref{fig: num-3pat} shows that the proposed Momentum-Net with SimpleCNN decreases RMSE dramatically in the first 30 layers, 
and tends to converge in 50 layers.
%Comparing the mean RMSE of Momentum-Net and WavResNet \MG{in Table~\ref{tab:numerical}, Momentum-Net with SimpleCNN obtained 4.5 HU RMSE reduction. } 
The Momentum-Net reduces the mean RMSE value by 4.5 HU and gives smaller standard deviations in RMSE, compared to WavResNet, as reported in Table~\ref{tab:numerical}.
This implies that the proposed Momentum-Net with SimpleCNN can improve both the accuracy and stability of low-dose CT image reconstruction than a state-of-the-art image denoising deep NN, WavResNet.
%Moreover, it \MG{gave} smaller standard deviations, meaning that it is more robust in reconstructing low-dose images. 
%With the SimpleCNN based Momentum-Net, we achieved convergent reconstruction results within 50 layers. 
The proposed Momentum-Net with SimpleCNN better removes noise and streak artifacts than WavResNet.
It also provides clearer reconstructions of some details; see, in Fig.~\ref{fig:mom-net-simplecnn}, the boundaries shown in the zoomed region at the top-right corner in the first example, the arrow pointed structures in zoomed areas of the second example, and the arrow pointed tissues in the left zoomed region in the third example.

%Visual results of three reconstructed test examples are shown in the second column of Fig.~\ref{fig:mom-net-simplecnn}. 

\subsection{Momentum-Nets involving RSN-based training}
We show the reconstructed examples by Momentum-Net with SimpleCNN-RSN and Dn-RSN in the third and fourth columns of Fig.~\ref{fig:mom-net-simplecnn} respectively. Comparing the first three and the last columns in Fig.~\ref{fig:mom-net-simplecnn}, we observe that Momentum-Net with {SimpleCNN-RSN} provides generally noisier reconstructions than WavResNet and Momentum-Net with SimpleCNN. 
However, Momentum-Net with SimpleCNN-RSN sometimes can provide clearer details than WavResNet. 
For example, in the right zoomed box of the second example, Momentum-Net with SimpleCNN-RSN shows better reconstruction quality for the arrow pointed structures than WavResNet, and in the left zoomed box in the third row, the former gives clearer small tissues marked by red arrows than the latter. 
Table~\ref{tab:numerical} reports that Momentum-Net with SimpleCNN-RSN is approximately 2.9 RMSE (HU) higher than WavResNet, while it has smaller standard deviations. This implies that Momentum-Net with SimpleCNN-RSN is more stable than WavResNet, although it may not provide better image qualities. Momentum-Net with Dn-RSN, however, provides the worst visual and numerical results among the compared four methods in this paper. 
%This is probably because in RSN-based learning, image denoisers without skip connection need higher NN complexity than residual mapping operators.

\section{Conclusion}\label{sec:conclusion}
We proposed an accurate and stable Momentum-Net architecture for LDCT image reconstruction. By applying a four-layer residual CNN to image refining modules, we achieved significant reconstruction improvements in both numerical and visual results, compared with a state-of-the-art noniterative LDCT denoising method WavResNet \cite{WavResNet18}. 
We additionally investigated how does learning nonexpansive mappings with the spectral normalization technique RSN affect Momentum-Net performance. 
Applying RSN to learning either residual mapping or image denoising NNs does not improve the reconstruction performance of Momentum-Net in low-dose CT. In the future, we will investigate other training techniques that imposes some mathematical conditions to image refining NNs. We are also interested in developing methods to further accelerate the Momentum-Net convergence in LDCT image reconstruction.
\section{Acknowledgment}
The authors thank Dr. Cynthia McCollough, the Mayo Clinic, the American Association of Physicists in Medicine, and the National Institute of Biomedical Imaging and Bioengineering for providing the Mayo Clinic data.

\section{References}
\bibliographystyle{IEEEbib}
\small \bibliography{refs}
%{\small \bibliography{refs}}

\end{document}